\documentclass[twocolumn]{aastex63}

\newcommand{\xmm} {{\it XMM-Newton}}
\newcommand{\chandra} {{\it Chandra}}
\newcommand{\nustar} {{\it NuSTAR}}
\newcommand{\swift} {{\it Swift}}
\newcommand{\erosita} {eROSITA}
\newcommand{\swiftxrt} {{\it Swift}/XRT}
\newcommand{\swiftuvot} {{\it Swift}/UVOT}
\newcommand{\gaia} {{\it Gaia}}

\newcommand{\cmsq} {cm$^{-2}$}
\newcommand{\nh} {$N_{\rm{H}}$}
\newcommand{\lx} {$L_{\rm{X}}$}
\newcommand{\fx} {$F_{\rm{X}}$}
\newcommand{\chisq} {$\chi^2$}

\newcommand{\oiii}{{\rm{[O\,\sc{iii}]}}}
\newcommand{\nii}{{\rm{[N\,\sc{ii}]}}}

\newcommand{\ha}{{\rm{H$\alpha$}}}
\newcommand{\hb}{{\rm{H$\beta$}}}
\newcommand{\hg}{{\rm{H$\gamma$}}}
\newcommand{\hd}{{\rm{H$\delta$}}}
\newcommand{\degree}{{$^\circ$}}
\newcommand{\ergs}{\mbox{\thinspace erg\thinspace s$^{-1}$}}
\newcommand{\ergcms}{\mbox{\thinspace erg\thinspace cm$^{-2}$\thinspace s$^{-1}$}}
\newcommand{\ergcmshz}{\mbox{\thinspace erg\thinspace cm$^{-2}$\thinspace s$^{-1}$ \thinspace Hz$^{-1}$}}
\newcommand{\ergshz}{\mbox{\thinspace erg\thinspace s$^{-1}$ \thinspace Hz$^{-1}$}}
\newcommand{\cntrt}{counts\,s$^{-1}$}
\newcommand{\kms}{km\,s$^{-1}$}

\newcommand{\mbh} {$M_{\rm BH}$}

\newcommand{\lamedd} {$\lambda_{\rm Edd}$}

\newcommand{\msol} {$M_{\odot}$}
\newcommand{\galaxy} {SDSS J143359.16+400636.0}

\newcommand{\nobsd}{27}

\shorttitle{An X-ray luminous TDE in SDSS J143359.16+400636.0}
\shortauthors{Brightman et al.}

\begin{document}

\title{A luminous X-ray transient in SDSS J143359.16+400636.0: a likely tidal disruption event}

\author{Murray Brightman}
\affiliation{Cahill Center for Astrophysics, California Institute of Technology, 1216 East California Boulevard, Pasadena, CA 91125, USA}

\author{Charlotte Ward}
\affiliation{Department of Astronomy, University of Maryland, College Park, MD 20742, USA}

\author{Daniel Stern}
\affiliation{Jet Propulsion Laboratory, California Institute of Technology, Pasadena, CA 91109, USA}

\author{Kunal Mooley}
\affiliation{Cahill Center for Astrophysics, California Institute of Technology, 1216 East California Boulevard, Pasadena, CA 91125, USA}

\author{Kishalay De}
\affiliation{Cahill Center for Astrophysics, California Institute of Technology, 1216 East California Boulevard, Pasadena, CA 91125, USA}

\author{Suvi Gezari}
\affiliation{Department of Astronomy, University of Maryland, College Park, MD 20742, USA}
\affiliation{Joint Space-Science Institute, University of Maryland, College Park, MD 20742, USA}

\author{Sjoert Van Velzen}
\affiliation{Department of Astronomy, University of Maryland, College Park, MD 20742, USA}
\affiliation{Center for Cosmology and Particle Physics, New York University, NY 10003, USA}

\author{Igor Andreoni}
\affiliation{Division of Physics, Mathematics and Astronomy, California Institute of Technology, Pasadena, CA 91125, USA}

\author{Matthew Graham}
\affiliation{Cahill Center for Astrophysics, California Institute of Technology, 1216 East California Boulevard, Pasadena, CA 91125, USA}

\author{Frank J. Masci}
\affiliation{IPAC, California Institute of Technology, 1200 East California Boulevard, Pasadena, CA 91125, USA}

\author{Reed Riddle}
\affiliation{Cahill Center for Astrophysics, California Institute of Technology, 1216 East California Boulevard, Pasadena, CA 91125, USA}

\author{Jeffry Zolkower}
\affiliation{Caltech Optical Observatories, California Institute of Technology, Pasadena, CA  91125, USA}

\email{murray@srl.caltech.edu}

\begin{abstract}
We present the discovery of a luminous X-ray transient, serendipitously detected by \swift's X-ray Telescope (XRT) on 2020 February 5, located in the nucleus of the galaxy \galaxy\ at $z=0.099$ (luminosity distance $D_{\rm L}=456$ Mpc). The transient was observed to reach a peak luminosity of $\sim10^{44}$ \ergs\ in the 0.3--10 keV X-ray band, which was $\sim20$ times more than the peak optical/UV luminosity. Optical, UV, and X-ray lightcurves from the Zwicky Transient Facility (ZTF) and \swift\ show a decline in flux from the source consistent with $t^{-5/3}$, and observations with \nustar\ and \chandra\ show a soft X-ray spectrum with photon index $\Gamma=2.9\pm0.1$. The X-ray/UV properties are inconsistent with well known active galactic nuclei (AGN) properties and have more in common with known X-ray tidal disruption events (TDE), leading us to conclude that it was likely a TDE. The broadband spectral energy distribution (SED) can be described well by a disk blackbody model with an inner disk temperature of $7.3^{+0.3}_{-0.8}\times10^{5}$ K, with a large fraction ($>40$\%) of the disk emission up-scattered into the X-ray band. An optical spectrum taken with Keck/LRIS after the X-ray detection reveals LINER line ratios in the host galaxy, suggesting low-level accretion on to the supermassive black hole prior to the event, but no broad lines or other indications of a TDE were seen. The stellar velocity dispersion implies the mass of the supermassive black hole powering the event is log(\mbh/\msol)$=7.41\pm0.41$, and we estimate that at peak the Eddington fraction of this event was $\sim$50\%. This likely TDE was not identified by wide-field optical surveys, nor optical spectroscopy, indicating that more events like this would be missed without wide-field UV or X-ray surveys.

\end{abstract}


\section{Introduction}

Tidal disruption events (TDEs) occur when stars in the center of a galaxy that orbit close to the supermassive black hole (SMBH) get close enough that the tidal forces acting on them exceed their own self gravity, causing the star to be disrupted. In this case a large fraction of the star's mass can be accreted onto the black hole producing a flare of electromagnetic radiation \citep[e.g.][]{rees88}. 

TDEs provide uniquely powerful tools for determining black hole demographics and investigating super-Eddington accretion. TDE rates are generally skewed to lower mass black holes, since the tidal disruption radius is interior of the Schwarzschild radius for \mbh$>10^{8}$ \msol, and therefore TDEs provide a useful signpost of lower mass SMBHs. Furthermore, for \mbh$<10^{7}$ \msol, TDEs can emit above the Eddington luminosity \citep{strubbe09}, making them laboratories for extreme accretion.

Distinguishing TDEs from flares of more common accretion onto an SMBH can be challenging \citep{auchettl18}. One defining feature of TDEs is that their luminosities decline monotonically, often with a power-law profile approximately following $t^{-5/3}$, determined by the time in which the stellar debris gets accreted \citep{evans89}.

While TDEs are regularly being discovered by wide-field optical surveys such as the Zwicky Transient Facility \citep[ZTF, e.g.][]{vanvelzen19} and the All-Sky Automated Survey for Supernovae \citep[ASAS-SN, e.g.][]{holoien19}, TDEs discovered in the X-rays are currently comparatively rare, although \erosita\ is set to change this, and has already identified a handful of candidate events \citep[e.g.][]{khabibullin20}. In general, optical/UV events have cooler spectra ($10^4$ K) and X-ray events have hotter ones ($10^5$ K) \citep{komossa15}.

We have recently begun a program to search through public \swiftxrt\ observations for transient sources. The {\it Neil Gehrels Swift Observatory} \citep[hereafter \swift,][]{burrows05} observes several tens of targets every day, many of which are monitoring observations with cadences of a few days, well suited to finding transient sources. With a field of view of 560 arcmin$^2$, \swiftxrt\ provides a great potential for serendipitously discovering X-ray transients in the fields of view of other targets \citep[e.g.][]{soderberg09}. Furthermore, since most \swift\ data are downloaded from the satellite and made public within hours of the observation, this allows the opportunity to follow up promptly in real time with other observatories.

On 2020 February 5, we serendipitously detected an X-ray source in the field of view (FoV) of a \swiftxrt\ observation of SN~2020bvc, a broad-lined Type Ic supernova in the galaxy UGC09379 \citep{ho20}, where no previous X-ray source had been detected. The position of the X-ray source was RA=14h 33m 58.96s, Decl.=+40\degree\ 06\arcsec\ 33.5\arcmin, with a positional uncertainty of 3.5\arcsec\ (90\% confidence). This is $\sim8$\arcmin\ from the supernova. The position of the X-ray source placed it in or near the galaxy \galaxy, different from SN~2020bvc. \galaxy\ has a spectroscopic redshift of $z=0.099$ (Section \ref{sec_keck}). Here we report on follow up and subsequent observations of the source which lead us to conclude that it was likely a X-ray TDE.

Throughout this paper we assume the cosmological parameters $H_{0}=70$~km\,s$^{-1}$\,Mpc$^{-1}$, $\Omega_{\rm m}=0.27$, and $\Omega_{\rm \Lambda}=0.73$. Under this assumed cosmology, the luminosity distance to \galaxy\ at $z=0.099$ is 456~Mpc. All uncertainties are quoted at the 90\% level unless otherwise stated.

\section{X-ray data analysis}
\subsection{Swift}
\label{sec_swift}

After the initial detection of the X-ray source, we requested follow up observations with \swift\ with both the XRT and Ultraviolet/Optical Telescope (UVOT) instruments, initially with a cadence of a few days, then a few times a month. In addition to the initial detection in the first XRT observation (obsID 00032818012), \swift\ has observed and detected the transient in the X-rays \nobsd\ times, all in photon counting mode. Previous to this, \swift\ observed the position of the source 17 times, 12 times in 2013 and 5 times in 2016 where the source was not detected in X-rays. We analyze all \swift\ observations here. 

\begin{figure}
\begin{center}
\includegraphics[width=90mm]{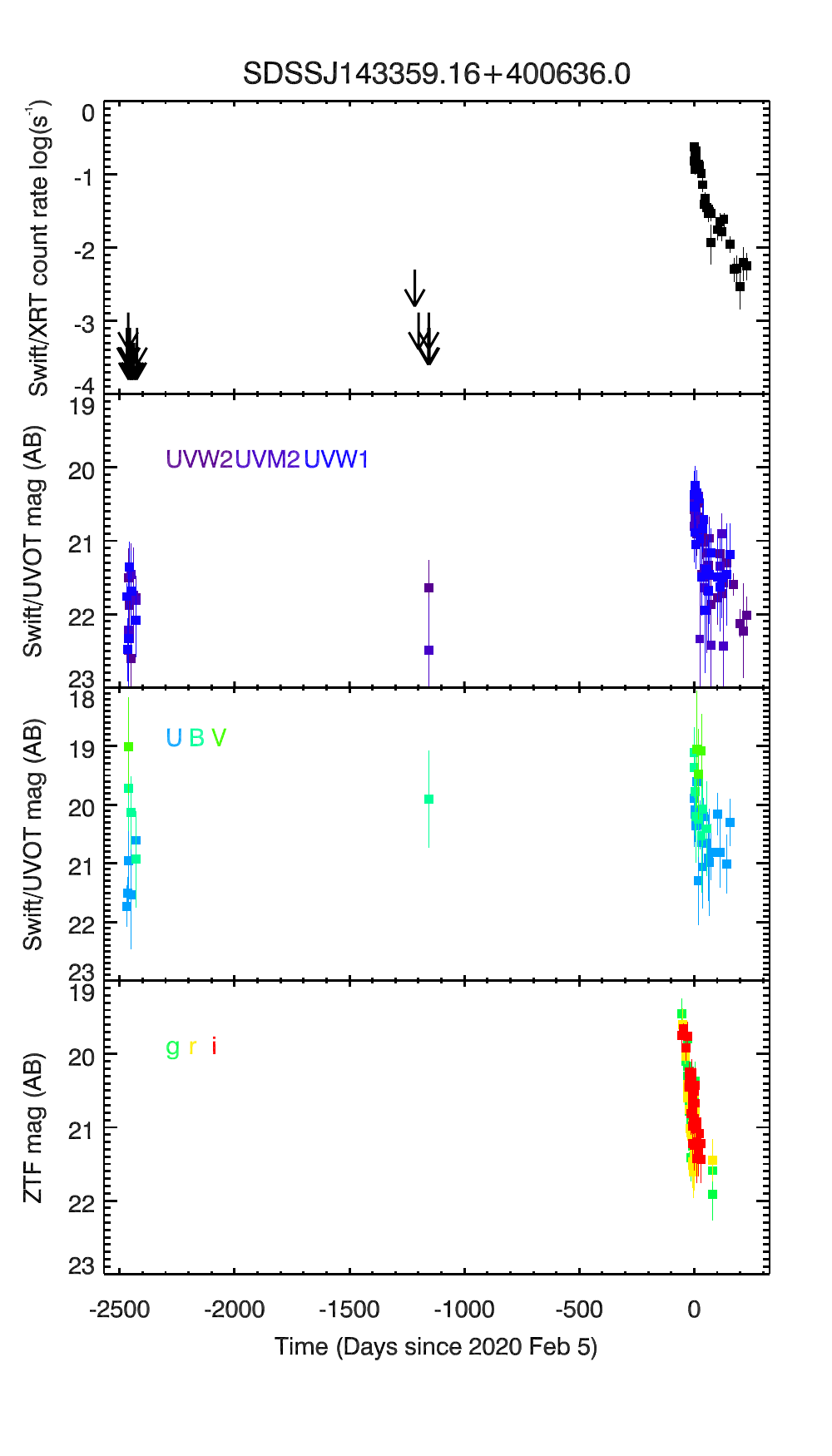}
\caption{Long-term lightcurve of \galaxy, from all \swiftxrt, \swiftuvot, and ZTF observations. On 2020 February 5 a bright X-ray source was detected with a count rate 2 orders of magnitude greater than previous upper limits (shown by downward pointing arrows). The host galaxy was seen in the UVOT data prior to the transient, so only a small increase in brightness was measured, and it was much less than seen in the X-rays. The ZTF data are from difference imaging, hence the host galaxy has been subtracted, and show the transient was detected in the optical $\sim60$ days before \swift\ detected it in the X-rays.}
\label{fig_ltcrv}
\end{center}
\end{figure}

In order to obtain an X-ray lightcurve of the source, we used the online tool provided by the University of Leicester\footnote{https://www.swift.ac.uk/user\_objects/} \citep{evans07,evans09}. All products from this tool are fully calibrated and corrected for effects such as pile-up and the bad columns on the CCD. The XRT lightcurve is shown in Figure \ref{fig_ltcrv}. For observations where a source has zero total counts, we estimate the 90\% upper limit on the count rate using a typical background count rate of 7$\times10^{-5}$ \cntrt\ and Poisson statistics. At peak, the transient event was detected at a brightness two orders of magnitude greater than these upper limits.


We also used the online tool as described above to build a stacked spectrum of the source. Furthermore, for each individual observation, we extracted events of the source using the {\sc heasoft} v6.25 tool {\sc xselect} \citep{arnaud96}. Source events were selected from a circular region with a 25\arcsec\ radius centered on the above coordinates, and a background region consisting of a larger circle external to the source region was used to extract background events. For each source spectrum, we constructed the auxiliary response file (ARF) using {\tt xrtmkarf}. The relevant response matrix file (RMF) from the CALDB was used. All spectra were grouped with a minimum of 1 count per bin.

The stacked spectrum has a total exposure time of 55 ks, and the average count rate of the source is (3.04$\pm0.08)\times10^{-2}$ \cntrt. We initially fitted the spectrum with an absorbed power-law model, {\tt tbabs*ztbabs*powerlaw} in {\sc xspec}, where the {\tt tbabs} model accounts for absorption in our Galaxy, fixed at 9.8$\times10^{19}$ \cmsq\ \citep{HI4PI16}, and {\tt ztbabs} accounts for absorption at the redshift of the source and is left as a free parameter. We find that \nh$=(7\pm3)\times10^{20}$ \cmsq, and the photon index $\Gamma=3.0\pm$0.2, where $C=252.93$ with 283 DoFs. We also tested a {\tt diskbb} model in place of the {\tt powerlaw} model, but it does not provide a good fit, where $C=489.66$ with 283 DoFs. However, the {\it addition} of a {diskbb} component to the {\tt tbabs*ztbabs*powerlaw} model does present a small improvement to the spectral fit, yielding \nh$=(1.1\pm0.1)\times10^{21}$ \cmsq, the temperature of the inner disk $kT=0.13^{+0.09}_{-0.03}$ keV, and photon index $\Gamma=2.8\pm$0.3, where $C=245.85$ with 283 DoFs.

\begin{figure}
\begin{center}
\includegraphics[width=90mm]{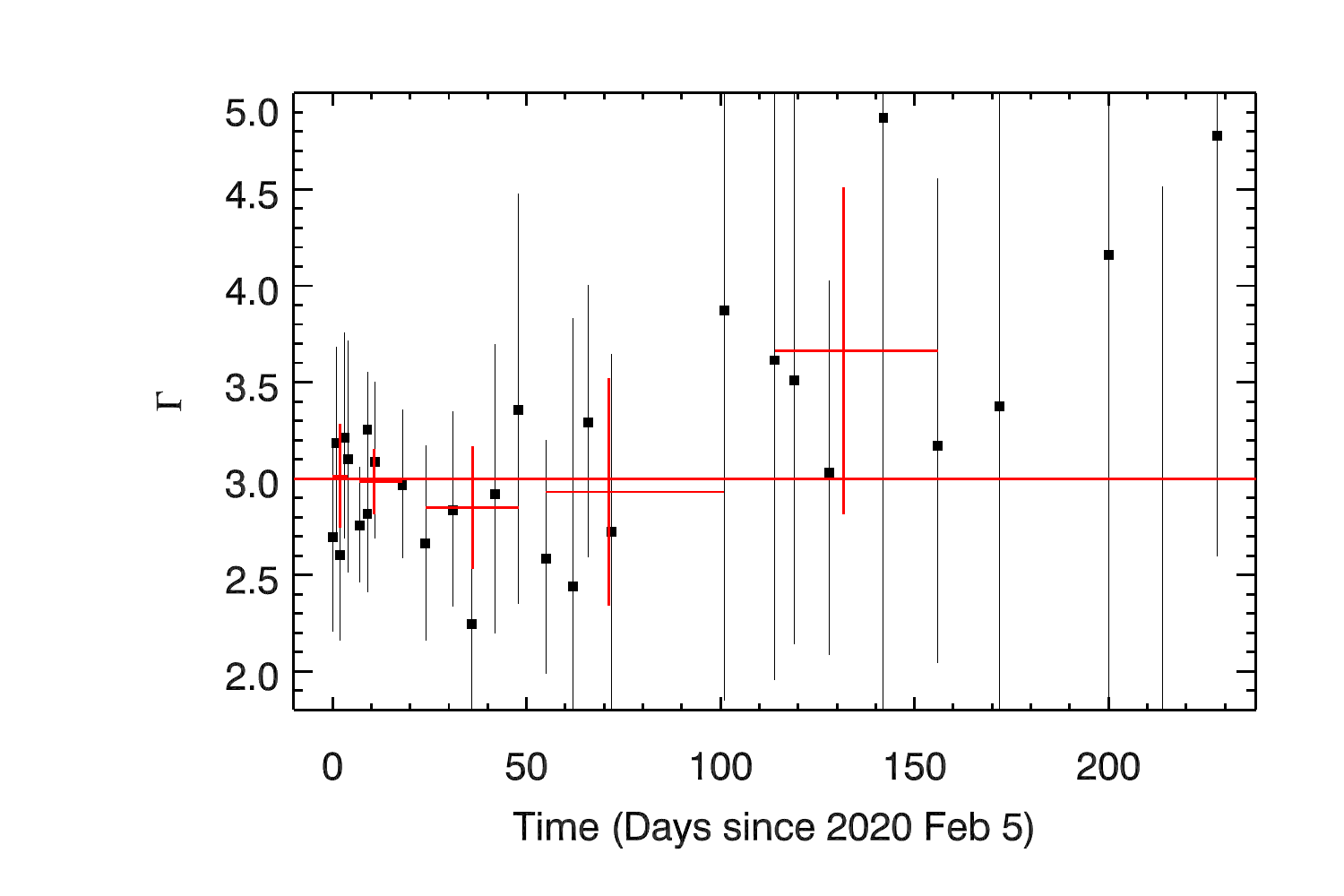}
\caption{The power-law index, $\Gamma$, of the fit to the \swiftxrt\ data as a function of time (black data points). The solid red line shows the value $\Gamma$ from the integrated spectrum. Also shown are binned averages where bins contain 5 observations each (red data points).}
\label{fig_gamma}
\end{center}
\end{figure}

Subsequently, we fitted the spectra from the individual obsIDs with the absorbed power-law model. We do not fit the more complicated model due to the low count nature of the individual spectra. Figure \ref{fig_gamma} shows the variation in $\Gamma$ over time, overplotted with binned averages (bins contain 5 observations each). There is no evidence of X-ray spectral evolution from the observations reported here. 

The observed (absorbed) 0.3--10 keV flux as measured by XRT is 5$\times10^{-12}$ \ergcms, which corresponds to a luminosity of 1$\times10^{44}$ \ergs at a distance of 456~Mpc. Assuming this model, the upper limit on the X-ray luminosity prior to the transient was $\sim10^{42}$ \ergs, corresponding to a $>2$ order-of-magnitude increase in the X-ray luminosity.

In addition to the XRT data, \swift\ also observed the source with its UVOT instrument, which has six filters, {\it UVW2} (central wavelength $\lambda=1928$ \AA), {\it UVM2} ($\lambda=2246$ \AA), {\it UVW1} ($\lambda=2600$ \AA), {\it U} ($\lambda=3465$ \AA), {\it B} ($\lambda=4392$ \AA), and {\it V} ($\lambda=5468$ \AA). In order to extract the photometry from the UVOT data, we used the tool {\tt uvotsource}, using circular regions with a 5\arcsec\ radius. Not every observation is taken with all six filters, however. We show the XRT and UVOT lightcurves in Figure \ref{fig_ltcrv}. While a UVOT source was clearly seen prior to the 2020 observations, likely emission from the host galaxy, a small increase in brightness measured by UVOT can be seen in the 2020 observations, though it is much weaker than seen in the X-rays. Also shown in Figure \ref{fig_ltcrv} are data from ZTF, which are described in Section \ref{sec_ztf}.

\subsection{NuSTAR}
\label{sec_nustar}

In order to study the hard X-ray emission from the transient, we obtained Director's Discretionary Time observation on the {\it Nuclear Spectroscopic Telescope Array} \citep[\nustar, obsID 90601606002,][]{harrison13}, which took place on 2020 February 13, 8 days after the X-ray transient was first detected by \swift. We used the {\sc heasoft} (v6.27) tool {\tt nuproducts} with default parameters to extract the \nustar\ spectrum. We used a circular region with a radius of 50\arcsec, centered on the peak of the emission to extract the source and a region with 100\arcsec\ radius to extract the background. The exposure time after filtering was 51.9~ks, from which the source was detected above background in each detector up to $\sim$15 keV, with a count rate of 0.01 \cntrt\ in the 3--15 keV band.

\subsection{Chandra}
\label{sec_chandra}

On 2020 February 16 and 29, 11 and 24 days after the initial \swift\ detection respectively, \galaxy\ was also serendipitously observed by \chandra\ \citep[obsIDs 23171 and 23172,][]{weisskopf99} for 10\,ks each exposure with ACIS-S at the aimpoint. These observations also targeted SN~2020bvc \citep{ho20}. This allowed us to obtain a better position of the source than \swiftxrt\ provided, and a higher signal-to-noise spectrum. 

In order to determine the position of the transient, we first ran the {\sc ciao} tool {\tt wavdetect} on the observations to obtain lists of positions for all sources in the \chandra\ FoV. Wavelet scales of 1, 2, 4, 8, and 16 pixels and a significance threshold of $10^{-5}$ were used. A total of 41 and 40 X-ray sources were detected in each observation, respectively.

We then cross-correlated the \chandra\ source lists with the {\it Gaia} DR2 catalog \citep{gaia18} to obtain the astrometric shifts. First we filtered to \gaia\ sources within 1\arcsec\ of the X-ray sources, excluding the transient itself, which left five \chandra/\gaia\ sources from both obsIDs. We define the astrometric shifts as the mean difference in RA and Dec between these matched sources. For obsID 23171, $\delta$RA$=-0.10\pm0.33$\arcsec\ and $\delta$Dec$=+0.55\pm0.28$\arcsec, and for obsID 23172, $\delta$RA$=+0.28\pm0.40$\arcsec\ and $\delta$Dec$=+0.01\pm0.38$\arcsec.

\begin{figure}
\begin{center}
\includegraphics[width=90mm]{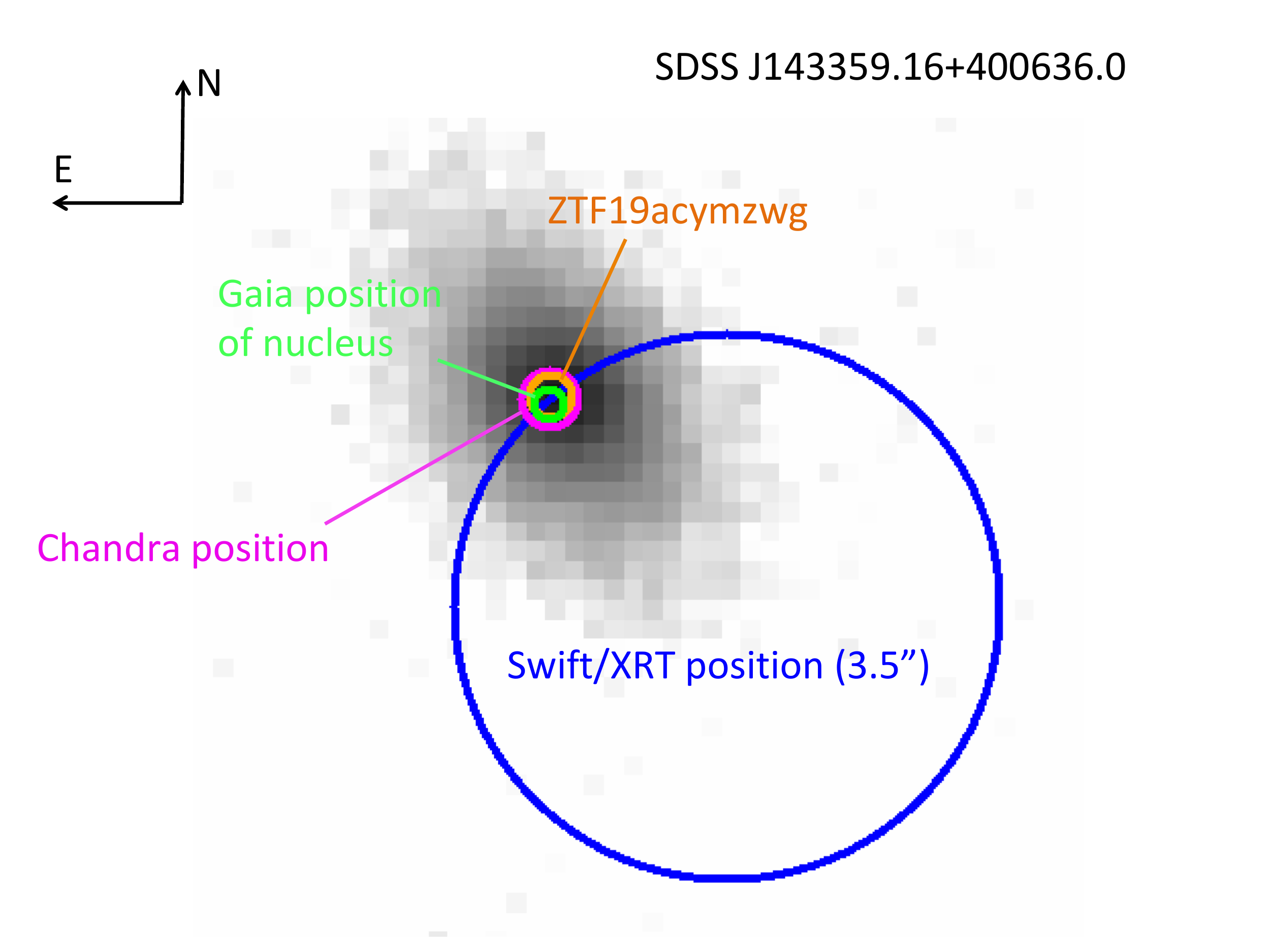}
\caption{PanSTARRS $i$-band image of the galaxy SDSS J143359.16+400636.0, where the green circle shows the \gaia\ position of the nucleus. The position of the X-ray transient detected by \swiftxrt\ is shown with a blue circle where the radius represents the 3.5\arcsec\ uncertainty (90\% confidence), which does not clearly place the source in the galaxy. The more accurate position provided by \chandra\ obsID 23172 is shown with a magenta circle (1$\sigma$ confidence), and identifies the transient with the nucleus of the galaxy. The orange circle shows the position of the related ZTF transient (1$\sigma$ confidence).}
\label{fig_img}
\end{center}
\end{figure}

Having applied the astrometric shifts to the \chandra\ source catalog, the position of the X-ray source from obsID 23171 is R.A. = 14h 33m 59.170s, Decl.=+40\degree\ 06\arcmin\ 36.18\arcsec\ (J2000), with an astrometric uncertainty of 0$\farcs$41 from the residual offsets with the {\it Gaia} catalog. From obsID 23172 the position is R.A. = 14h 33m 59.170s, Decl.=+40\degree\ 06\arcmin\ 36.10\arcsec\ (J2000), with an astrometric uncertainty of 0$\farcs$37 from the residual offsets with the {\it Gaia} catalog. The \gaia\ position of the nucleus is R.A. = 14h 33m 59.170s, Decl.=+40\degree\ 06\arcmin\ 36.05\arcsec\ (J2000). Figure \ref{fig_img} shows the PanSTARRS image of \galaxy, with the \gaia\ position of the nucleus shown with respect to the \chandra\ position of the X-ray source, which is coincident.

Also shown in Figure \ref{fig_img} is the position of ZTF19acymzwg, a candidate optical transient source detected in the $g$, $r$, and $i$ bands by the ZTF on 2019 December 14, 53 days prior to the detection of the X-ray transient by \swiftxrt. We describe the analysis of the ZTF data fully in Section \ref{sec_ztf}, including an updated position for the transient of RA=14h 33m 59.17s and Dec=+40\degree\ 06\arcmin\ 36.1\arcsec\ with a 1$\sigma$ positional uncertainty of 0.29\arcsec. ZTF19acymzwg is likely related to the X-ray transient one since their positions consistent with each other within the uncertainties.

Due to the relatively high count rate and readout time of the ACIS detectors, we check for pileup of the source using the  {\sc ciao} v4.11 tool {\sc pileup\_map}. We find that the pileup fraction is only $\sim2$ \% and therefore negligible. We then proceed to extract the spectrum of the source from both obsIDs, using the {\sc ciao}  tool {\sc specextract}, and an elliptical region with a semi-major axis of 7.7\arcsec\ and a semi-minor axis of 4.4\arcsec. We used this shape and size due to the source being off axis where the PSF is larger and elongated. Background events were extracted from a nearby region. The source was detected in the $\sim$10 ks observations with a count rate of $1.45\pm0.03\times10^{-1}$ \cntrt\ and $7.9\pm0.3\times10^{-2}$ \cntrt\ respectively in the 0.5--8 keV band in the ACIS-S detector. There is clear evidence for a drop in flux over the 13-day period between \chandra\ observations. Intra-observational lightcurves of the \chandra\ observations were also extracted, binning on various time scales, though none of these showed significant count rate variability during the observations. 

\begin{figure}
\begin{center}
\includegraphics[width=90mm]{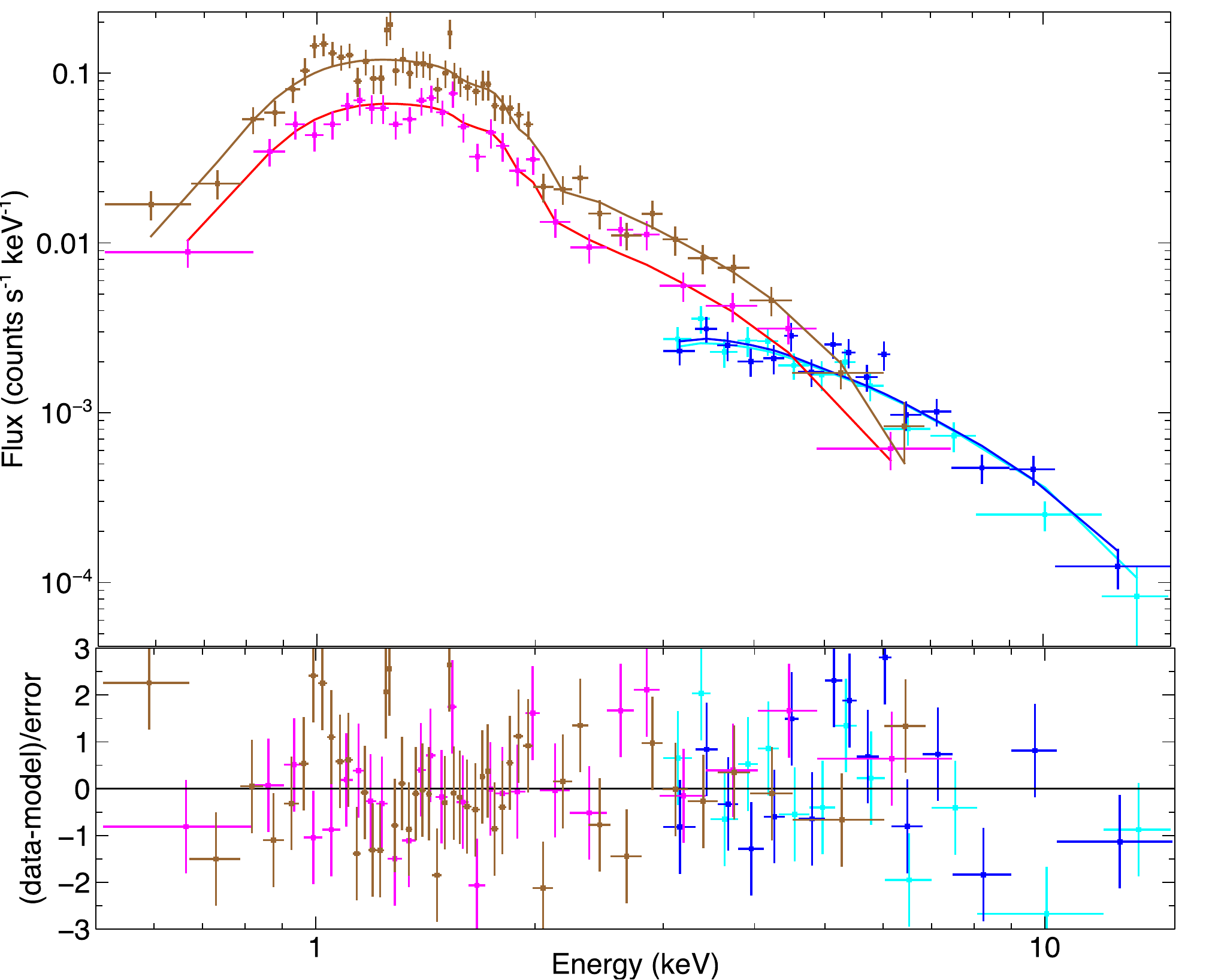}
\caption{\chandra\ obsID 23171 (brown), obsID 23172 (magenta), \nustar\ FPMA (blue) and FPMB (cyan) spectra of the X-ray transient in \galaxy, taken 8--24 days after the \swiftxrt\ detection. The data are consistent with an absorbed power-law, with a constant to account for flux variability between data sets, plotted here as solid lines. The data have been binned for plotting clarity, where each bin has a minimum $5\sigma$ detection.}
\label{fig_spec}
\end{center}
\end{figure}

We jointly fitted the \nustar\ spectra with both \chandra\ spectra in {\sc xspec} using the C-statistic and a cross-calibration constant included to account for cross-calibration uncertainties and flux variability. The spectra are plotted in Figure \ref{fig_spec}, which shows that they are well described by a simple absorbed power-law ({\tt constant*zTbabs*powerlaw}) over the 0.5--15 keV range, with \nh$=(9\pm5)\times10^{20}$ \cmsq\ and $\Gamma=2.9\pm0.1$, where $C=847.66$ with 883 DoFs, consistent with the integrated \swift\ spectrum. The absorption measured is in excess of the Galactic value 9.8$\times10^{19}$ \cmsq\ and is therefore attributable to the host. 

As with the integrated \swift\ spectrum in Section \ref{sec_swift}, we fitted other spectral models to the joint \chandra\ and \nustar\ spectra. Again a fit with a {\tt diskbb} model instead of a {\tt powerlaw} model does not fit the spectrum well, with $C=1371.92$ with 883 DoFs. The addition of a {\tt diskbb} model to the {\tt powerlaw} model does not produce an improvement to the fit, where $C=847.12$ with 881 DoFs, at odds with the \swift\ data. Since the \swift\ data were integrated over all exposures, which covered a larger and later time span than the Chandra data, it's possible that this {\tt diskbb} component emerged at later times. We checked this by creating a \swift\ spectrum which covered the same time period as the \chandra\ observations. As with the full \swift\ dataset, the addition of the {\tt diskbb} model to the {\tt powerlaw} model produces an improvement to the fit, arguing against the above hypothesis and leaving the \chandra\ and \swift\ data at odds with each other. Since the \chandra\ data have a higher number of counts with higher signal-to-noise than the \swift\ data, we defer to the \chandra\ results, concluding that there is no evidence for a {\tt diskbb} component in addition to the powerlaw one.

The cross-calibration constant for \nustar\ FPMA, $C_{\rm FPMA}$, is fixed to unity, while $C_{\rm FPMB}$ is fixed to 1.04 \citep{madsen15}. The constants for \chandra\ are $C_{\rm 23171}=1.36^{+0.18}_{-0.16}$ and $C_{\rm 23172}=0.91\pm0.05$. The 0.5--15 keV flux, as measured 2020 February 13, 8 days after the X-ray transient was first detected by \swift, is 4.0$\times10^{-12}$ \ergcms, which corresponds to a luminosity of 9.8$\times10^{43}$ \ergs\ at a distance of 456~Mpc. These X-ray spectral fitting results are summarized in Table \ref{table_fitxray}.

\begin{table}
\centering
\caption{\nustar\ and \chandra\ X-ray spectral fitting results.}
\label{table_fitxray}
\begin{center}
\begin{tabular}{l l l l l}
\hline
Parameter & Result \\
\hline
\nh & $(9\pm5)\times10^{20}$ \cmsq\ \\
$\Gamma$ & $2.9\pm0.1$ \\
Normalization & $(1.2\pm0.2)\times10^{-3}$ \\
\fx\ (0.5--15 keV) & $4.0^{+0.2}_{-0.4}\times10^{-12}$ \ergcms\ \\
\lx\ (0.5--15 keV) & $9.8^{+0.2}_{-0.4}\times10^{43}$ \ergs\ \\
$C_{\rm FPMA}$ & 1.0 (fixed) \\
$C_{\rm FPMB}$ & 1.04 (fixed) \\
$C_{\rm 23171}$ & $1.36^{+0.18}_{-0.16}$ \\
$C_{\rm 23172}$ & $0.91\pm0.05$ \\
C-statistic & 847.66 \\
DoFs & 883 \\ 
\hline
\end{tabular}
\tablecomments{Results from the fit of an absorbed powerlaw to the \nustar\ and \chandra\ spectra of the X-ray transient in \galaxy\ as measured 2020 February 13, 8 days after the X-ray transient was first detected by \swift.}
\end{center}
\end{table}

\subsection{eROSITA}
\cite{khabibullin20} reported via The Astronomer's Telegram (\#13494) the detection by {\it Spectrum-Roentgen-Gamma} ({\it SRG})/eROSITA of a very bright X-ray source, SRGet J143359.25+400638.5, centered on \galaxy\ on 2019 December 27, 40 days prior to the detection of the transient with \swiftxrt. The reported 0.3--8 keV flux was $6.5\times10^{-12}$ \ergcms, with no reported variability over the 11 individual scans with an interval of 4 hours. This reported flux is almost the same flux that \swiftxrt\ measured, suggesting that the X-ray flux of the source remained approximately constant for at least 40 days prior to the detection by \swiftxrt, before declining, or rose and fell, or vice versa. The X-ray spectrum was reported to be soft and described by a disk black body spectrum with a temperature of 0.29 keV. We simulate a spectrum with these model parameters and fit with a power-law model, which yields $\Gamma=2.9$, which is the same as measured by \swiftxrt, indicating that no spectral evolution took place between the \erosita\ detection and the \swiftxrt\ one. The authors suggested an association with ZTF19acymzwg which we confirm here.

\subsection{ROSAT}
{\galaxy\ was not detected by the {\it ROentgen SATellite} ({\it ROSAT}) in its all-sky survey performed in 1990 \citep[RASS,][]{voges99,boller16}. We calculate an X-ray flux upper limit for \galaxy\ from the {\it ROSAT} data using the {\tt SOSTA} (source statistics) tool available in the {\sc heasoft} {\tt XIMAGE} image processing package. The 3-$\sigma$ upper limit on the 0.1--2.4 keV count rate calculated using this method is 0.07 \cntrt. Assuming the spectral shape measured by \nustar\ and \chandra\ above, this corresponds to a 0.3--10 keV flux of 1.1$\times10^{-12}$ \ergcms, which is slightly less than the peak flux of the event. This limit is also similar to that of the XMM Slew Survey \citep{saxton12a}.}

\section{Zwicky Transient Facility}
\label{sec_ztf}

ZTF is an optical time-domain survey that uses the Palomar 48-inch Schmidt telescope with a 48 deg$^2$ field of view and scans more than 3750 deg$^2$ an hour to a depth of 20.5 mag \citep{bellm19,graham19,masci19}. As described in Section \ref{sec_chandra}, the candidate optical transient ZTF19acymzwg was detected in the $g$, $r$, and $i$ bands by ZTF on 2019 December 14, 53 days before the detection with \swiftxrt. Previous to this date, the field was observed on 2019 October 5 and the transient was not detected in any filter. 

First, in order to determine the position of the transient we use \texttt{The Tractor} \citep{lang16} to forward model the host galaxy profile and the transient point source position. \texttt{The Tractor} forward models in pixel space by parametrizing a sky noise and point spread function model for each image and modeling this simultaneously with each source's shape, flux and position. We apply the modeling to 49 $g$, $r$ and $i$-band ZTF images with limiting magnitude $> 21.5$ taken from 2019 December 29 to 2020 March 28 when the transient is bright in these bands. We find that the galaxy is better modeled by a de Vaucoleur profile than an exponential profile and that the transient point source position is given by RA=14h 33m 59.17s and Dec=+40\degree\ 06\arcmin\ 36.1\arcsec\ with a 3$\sigma$ positional uncertainty of 0.61\arcsec.

Once we obtained the position of the transient, we produced ZTF lightcurves using the ZTF forced-photometry service \citep{masci19} to produce difference-imaging photometry at the best-fit transient position across all ZTF images of the field taken between 2018 March 21 and 2020 May 11. We found no evidence for nuclear activity before the flare. The ZTF difference magnitudes are plotted in Figure \ref{fig_ltcrv}, along with the \swift\ lightcurve.

\section{Keck/LRIS optical spectroscopy}
\label{sec_keck}

\begin{figure}
\begin{center}
\includegraphics[width=90mm]{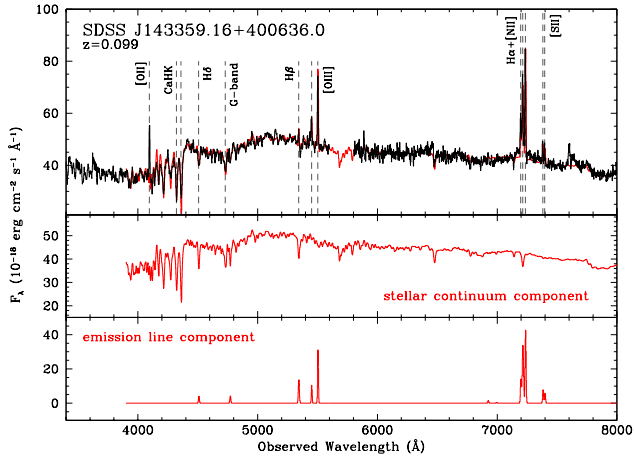}
\caption{Keck/LRIS spectrum of the nucleus of \galaxy\ (top) taken on 2020 February 18 (black), 13 days after the X-ray transient was detected by \swift. Key emission lines are labelled.  The model fit to the spectrum is underplotted (red), consisting of a stellar continuum component (middle) and an emission line component (bottom).}
\label{fig_keck_spec}
\end{center}
\end{figure}

We obtained an optical spectrum of the host galaxy nucleus with Keck/LRIS \citep{oke95} on 2020 February 18, 13 days after the initial \swift\ detection. The data were acquired using a standard long slit mode using a 1\arcsec\ slit on both the red and blue sides when the seeing was 1.01\arcsec\ in $i$ band. The spectra were reduced using standard long slit reduction procedures, including flat-fielding, wavelength calibration using arcs and flux calibration using a standard star as implemented in the \texttt{lpipe} package \citep{perley19}. The spectrum in shown in Figure \ref{fig_keck_spec}. 

We proceeded to fit the Keck/LRIS spectrum in order to determine the velocity dispersion from the stellar absorption lines and the fluxes of the emission lines. We applied Penalized Pixel-Fitting \citep{cappellari04,cappellari17} to the spectrum which finds the velocity dispersion of stellar absorption lines using a large sample of high spectral resolution templates of single stellar populations adjusted to match the resolution of the input spectrum. We simultaneously fitted the narrow H$\alpha$, H$\beta$, H$\gamma$, H$\delta$,  [S\,{\sc ii}] 6717, 6731, [N\,{\sc ii}] 6550, 6585, [O\,{\sc i}] 6302, 6366 and [O\,{\sc iii}] 5007, 4959 emission lines during template fitting. The emission line fluxes were each fit as free parameters but the line widths of the Balmer series were tied to each other, as were the line widths of the forbidden lines. We show the best fit model to the Keck/LRIS spectrum, including both the emission line component and the stellar continuum component, in Figure \ref{fig_keck_spec}. equivalent widths of the lines are presented in Table \ref{table_ews}. The The redshift of the galaxy was also determined to be 0.099.

\begin{table}
\centering
\caption{Equivalent widths of the narrow lines observed in the Keck/LRIS spectrum.}
\label{table_ews}
\begin{center}
\begin{tabular}{l l l}
\hline
Line & EW (\AA) & EW error (\AA) \\
\hline
\hd & 0.832 &  0.257 \\
\hg &  0.793 &  0.209 \\
\hb &  2.575 &  0.429 \\
\ha &  8.627 &  0.365 \\
{\rm{[S\,\sc{ii}]}}6716 & 1.523 &  0.208 \\
{\rm{[S\,\sc{ii}]}}6731 & 1.125 &  0.213 \\
\oiii5007 &  6.009 &  0.408 \\
{\rm{[O\,\sc{i}]}}6300 &  0.363 &  0.230 \\
\nii6583 &  10.286 &  0.409 \\
\hline
\end{tabular}
\end{center}
\end{table}

The velocity dispersion of the stellar absorption lines was determined to be $213\pm12$ \kms. We used this to calculate the black hole mass from the \mbh-$\sigma_{*}$ relation, using the fit to the reverberation-mapped AGN sample from \cite{woo13}, and the following formula, log(\mbh/\msol) = $\alpha + \beta $log$(\sigma_{*}/200$ \kms), where $\alpha =7.31\pm0.15$ and $\beta = 3.46\pm0.61$. The intrinsic scatter of this relation is $\epsilon = 0.41\pm0.05$. This yielded log(\mbh/\msol)$=7.41\pm0.41$. Alternatively, if we use the quiescent galaxy + AGN sample from \cite{woo13}, where $\alpha =8.36\pm0.05$ and $\beta = 4.93\pm0.28$ with intrinsic scatter $\epsilon = 0.43\pm0.04$, the black hole mass estimate is log(\mbh/\msol)$=8.49\pm0.43$, an order of magnitude more massive.

We then plotted the emission line ratios \oiii/\hb\ and \nii/\ha\ in Figure \ref{fig_bpt}, along with the diagnostic lines from \cite{kewley01} and {\cite{kauffmann03} to determine the excitation mechanism of the narrow lines. The line ratios place the nucleus of \galaxy\ in the LINER region of this diagnostic diagram, almost at the border of the Seyfert region. The stellar absorption template fitting predicted strong \hb\ absorption, which is why we see a high \oiii/\hb\ ratio in the initial spectrum. The lack of broad lines classifies the nucleus as a type 2 LINER. Also plotted on Figure \ref{fig_bpt} are the line ratios of nine optically- and radio-selected TDE hosts \citep{lawsmith17,french16,french17,mattila18,anderson19} along with SDSS galaxies for comparison. The TDE-host and galaxy emission line flux data have been taken from the SDSS DR7 MPI-JHU catalog\footnote{https://wwwmpa.mpa-garching.mpg.de/SDSS/DR7/}, where the stellar absorption-line spectra have also been subtracted before measurement \citep{kauffmann03,tremonti04}.

\begin{figure}
\begin{center}
\includegraphics[width=90mm]{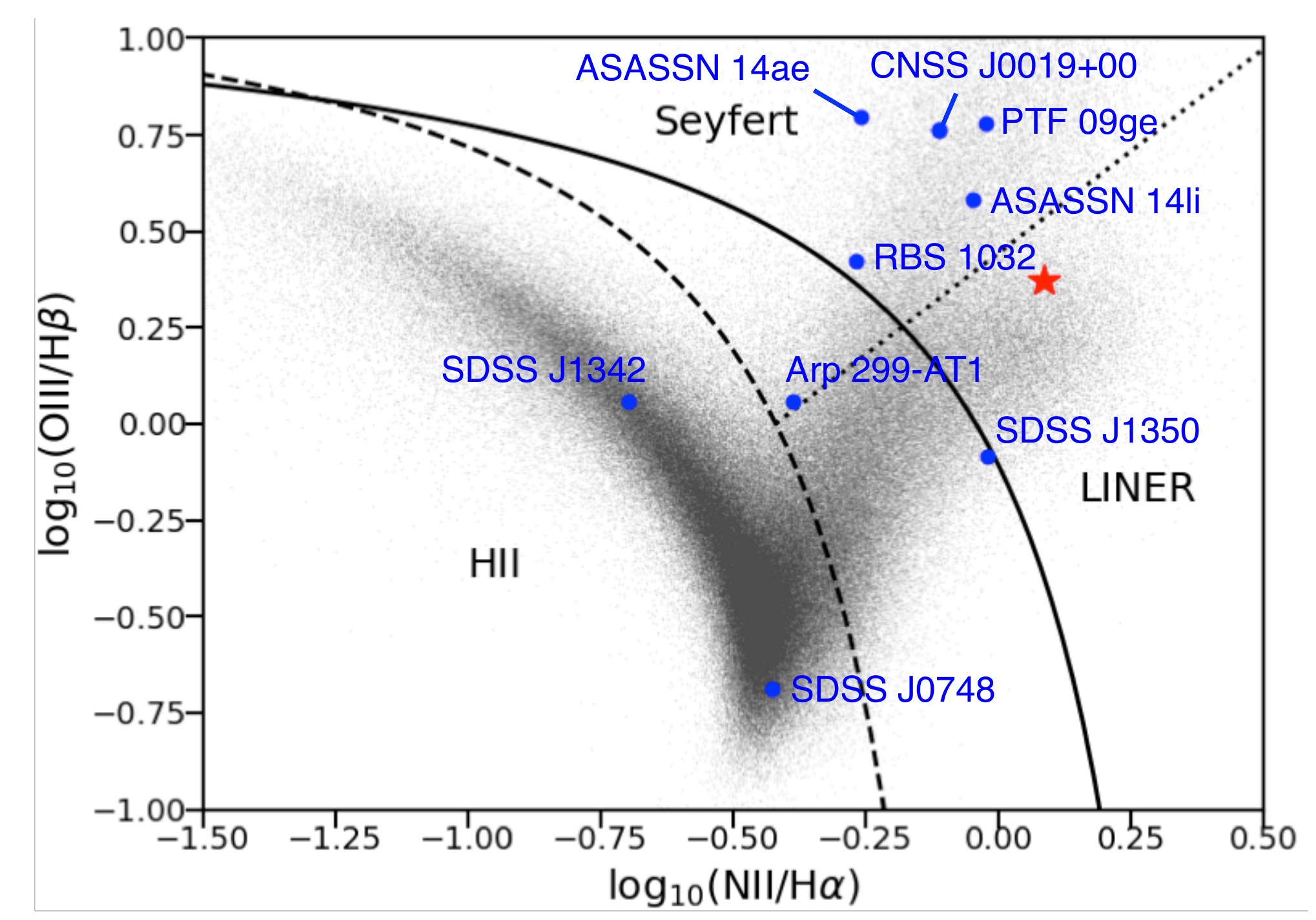}
\caption{Emission line ratio diagnostic diagram showing where \galaxy\ (red star) lies with respect to the Seyfert, LINER and star-forming (HII) regions. The nucleus lies in the LINER region, indicating that AGN activity was present, at least at a low level, before the onset of the transient. For comparison, data from SDSS on optically- and radio-selected TDE hosts are shown as blue circles and labelled, and other galaxies are shown in gray, where the stellar absorption-line spectrum has been subtracted.}
\label{fig_bpt}
\end{center}
\end{figure}

Since the narrow lines are produced in the narrow line region, which can be kiloparsecs from the SMBH \citep[e.g.][]{chen19}, this tells us that \galaxy\ had low-level AGN activity some time before the onset of the X-ray transient. To determine the spatial extent of the narrow line region in \galaxy, we analyzed the 2-dimensional Keck/LRIS spectrum. This shows the galaxy emission had a spatial extent of $\sim$5.2\arcsec. We extracted a spectrum from each edge of the galaxy which were separated by a 2.4\arcsec\ gap and each extraction region had a width of 1.4\arcsec. In both edge spectra, we located narrow line emission from the \oiii\ 5007\AA, 4959\AA\ doublet. This suggests that the narrow line emission has a spatial extent of $\sim2.4$\arcsec. Given the scale of 1.831 kpc/\arcsec\ at this redshift under our assumed cosmology, this implies the narrow lines were produced at a projected distance of 4.4~kpc, and that they were illuminated at least 10,000 years prior to the transient.

The flux of the \oiii\ line is $3.78\pm0.15\times10^{-16}$ \ergcms. From an investigation of the relationship between X-ray and optical line emission in 340 \swift/BAT-selected AGN \citep{berney15}, the \oiii\ flux expected from the 2--10 keV flux of $10^{-12}$ \ergcms, the peak X-ray flux measured by \swiftxrt, is in the range of $10^{-15}$--$10^{-13}$ \ergcms, higher than what we measure. The lower than expected \oiii\ flux we measure indicates that the AGN was at a low luminosity prior to the transient. This is also consistent with the upper limits on the X-ray luminosity of the nucleus prior to the transient, which at $\sim10^{42}$ \ergs, is relatively low for an AGN.

\section{Karl G. Jansky Very Large Array}
\label{sec_vla}

We carried out radio observations with the Karl G. Jansky Very Large Array (VLA) through Director's Discretionary Time (project code VLA/20A-579, PI: Mooley) on 2020 August 2, 180 days after the detection by \swift. Data were obtained at C band in the 3-bit mode of the WIDAR correlator to get a contiguous frequency coverage between 4--8 GHz. Standard VLA calibrator sources 3C286 and J1416+3444 were used to calibrate the flux/bandpass and phases respectively. The data were processed using the NRAO CASA pipeline and imaged using the {\tt clean} task in CASA. 

We did not detect any radio source at the location of the transient, and place a 3$\sigma$ upper limit of 28~$\mu$Jy on the 6 GHz flux density. We can therefore place an upper limit of $4 \times 10^{37}$\,erg\,s$^{-1}$ on the radio luminosity at a distance of 456~Mpc. The closest X-ray observation in time to the VLA one was by \swiftxrt\ on 2020 July 27 (obsID 00013265017), where we measured a 0.3--10 keV luminosity of $3.8\times10^{42}$ \ergs. The X-ray-to-radio luminosity ratio is therefore $>10^5$. Comparing our radio upper limit with the radio emission seen in jetted TDEs \citep[e.g.][]{alexander2020}, we can rule out the presence of a relativistic jet.

\section{Lightcurve fitting}
\label{sec_ltcrv}

After the initial detection by \swift, the lightcurve of the transient appeared to monotonically decline in flux, shown by \swift, \nustar, \chandra, and ZTF. In order to infer more details regarding the nature of the source, we fitted the lightcurve of the source in each band with a power-law model, $F=A(t-t_{\rm 0})^n+C$, where $F$ is the observed flux density of the source, $A$ is a normalization constant, $t$ is the time in days since the transient was first detected by \swift\ (2020 February 5), $t_{\rm 0}$ is the inferred start time of the event in days, and $n$ is the power-law index. $C$ is a constant which represents the underlying emission from the galaxy in UVOT data only, and set to zero for the XRT data since no X-ray emission is seen from the galaxy, and set to zero for the ZTF data since the galaxy has already been subtracted in these data. We determine the underlying emission from the galaxy in the UVOT data by averaging over the photometry measured previous to the detection of the transient. 

We calculate the 2 keV monochromatic flux density as measured from the power-law model fit to the \nustar\ and \chandra\ data with the \nh\ and $\Gamma$ parameters fixed to their best-fit values. We use the UVOT flux densities as produced by {\tt uvotsource}, and the ZTF difference imaging fluxes. We present these measurements in Tables \ref{table_fluxes} and \ref{table_ztf}.

\begin{table*}
\centering
\caption{\swift/XRT and UVOT fluxes.}
\label{table_fluxes}
\begin{center}
\begin{tabular}{c c c c c c c c c}
\hline
Time & obsID & 2 keV & UVW2 & UVM2 & UVW1 & U & B & V\\
(Days) &  & ($\mu Jy$) & ($mJy$) & ($mJy$) & ($mJy$) & ($mJy$) & ($mJy$) & ($mJy$)\\
\hline
           0&00032818012 & 0.260$\pm$ 0.084& 0.023$\pm$ 0.004& 0.024$\pm$ 0.004& 0.023$\pm$ 0.006& 0.040$\pm$ 0.016& 0.065$\pm$ 0.031& 0.004$\pm$ 0.063\\
           1&00032818014 & 0.261$\pm$ 0.098& 0.021$\pm$ 0.004& 0.017$\pm$ 0.004& 0.026$\pm$ 0.008& 0.012$\pm$ 0.014& 0.082$\pm$ 0.033& -\\
           2&00032818015 & 0.340$\pm$ 0.098& 0.022$\pm$ 0.004& 0.026$\pm$ 0.005& 0.016$\pm$ 0.006& 0.031$\pm$ 0.016& 0.015$\pm$ 0.030& 0.029$\pm$ 0.068\\
           3&00032818016 & 0.208$\pm$ 0.084& 0.020$\pm$ 0.005& 0.020$\pm$ 0.005& 0.022$\pm$ 0.007& 0.034$\pm$ 0.017& -& -\\
           4&00032818017 & 0.145$\pm$ 0.064& 0.021$\pm$ 0.004& 0.019$\pm$ 0.004& 0.030$\pm$ 0.007& 0.045$\pm$ 0.017& 0.045$\pm$ 0.031& 0.070$\pm$ 0.071\\
           7&00032818018 & 0.333$\pm$ 0.068& 0.017$\pm$ 0.003& 0.024$\pm$ 0.004& 0.014$\pm$ 0.004& 0.026$\pm$ 0.011& 0.030$\pm$ 0.022& 0.034$\pm$ 0.060\\
           9&00089025001 & 0.182$\pm$ 0.042& 0.017$\pm$ 0.001& -& -& -& -& -\\
           9&00032818019 & 0.233$\pm$ 0.066& -& -& -& -& -& -\\
          11&00032818020 & 0.187$\pm$ 0.055& 0.015$\pm$ 0.004& 0.017$\pm$ 0.005& 0.027$\pm$ 0.008& 0.052$\pm$ 0.019& 0.027$\pm$ 0.032& 0.087$\pm$ 0.080\\
          18&00013265001 & 0.208$\pm$ 0.057& 0.019$\pm$ 0.003& 0.025$\pm$ 0.005& 0.025$\pm$ 0.005& 0.011$\pm$ 0.008& 0.029$\pm$ 0.017& 0.058$\pm$ 0.042\\
          24&00013265002 & 0.215$\pm$ 0.072& 0.022$\pm$ 0.004& 0.004$\pm$ 0.003& 0.015$\pm$ 0.005& 0.027$\pm$ 0.010& 0.011$\pm$ 0.021& -\\
          31&00013265003 & 0.159$\pm$ 0.056& 0.009$\pm$ 0.002& 0.002$\pm$ 0.003& 0.009$\pm$ 0.005& 0.020$\pm$ 0.009& 0.022$\pm$ 0.020& 0.084$\pm$ 0.049\\
          36&00013265004 & 0.163$\pm$ 0.077& 0.015$\pm$ 0.004& -& 0.017$\pm$ 0.005& 0.014$\pm$ 0.009& 0.034$\pm$ 0.020& -\\
          42&00013265005 & 0.055$\pm$ 0.028& 0.014$\pm$ 0.002& 0.010$\pm$ 0.003& 0.019$\pm$ 0.005& 0.030$\pm$ 0.009& 0.009$\pm$ 0.016& -\\
          48&00013265006 & 0.049$\pm$ 0.037& 0.009$\pm$ 0.003& 0.008$\pm$ 0.005& 0.006$\pm$ 0.005& 0.003$\pm$ 0.010& -& 0.022$\pm$ 0.053\\
          55&00013265007 & 0.067$\pm$ 0.027& 0.012$\pm$ 0.003& 0.006$\pm$ 0.003& 0.010$\pm$ 0.004& 0.020$\pm$ 0.009& 0.025$\pm$ 0.019& -\\
          62&00013265008 & 0.061$\pm$ 0.048& 0.010$\pm$ 0.003& 0.008$\pm$ 0.005& 0.008$\pm$ 0.005& 0.016$\pm$ 0.011& 0.009$\pm$ 0.022& 0.026$\pm$ 0.058\\
          66&00013265009 & 0.038$\pm$ 0.021& 0.014$\pm$ 0.004& 0.009$\pm$ 0.004& 0.010$\pm$ 0.006& 0.015$\pm$ 0.012& -& -\\
          72&00013265010 & 0.045$\pm$ 0.028& 0.006$\pm$ 0.002& 0.004$\pm$ 0.003& 0.013$\pm$ 0.004& 0.017$\pm$ 0.008& 0.010$\pm$ 0.021& 0.002$\pm$ 0.039\\
         101&00013265011 & 0.011$\pm$ 0.026& 0.007$\pm$ 0.002& -& 0.009$\pm$ 0.005& 0.032$\pm$ 0.011& -& 0.009$\pm$ 0.045\\
         114&00013265012 & 0.024$\pm$ 0.038& 0.012$\pm$ 0.003& 0.010$\pm$ 0.004& 0.008$\pm$ 0.005& 0.017$\pm$ 0.010& 0.002$\pm$ 0.018& 0.001$\pm$ 0.044\\
         119&00013265013 & 0.018$\pm$ 0.018& 0.007$\pm$ 0.002& 0.016$\pm$ 0.004& 0.009$\pm$ 0.005& 0.007$\pm$ 0.009& -& -\\
         128&00013265014 & 0.029$\pm$ 0.020& 0.008$\pm$ 0.002& 0.004$\pm$ 0.003& 0.005$\pm$ 0.007& -& 0.023$\pm$ 0.023& -\\
         142&00013265015 & 0.001$\pm$ 0.000& 0.011$\pm$ 0.002& -& 0.010$\pm$ 0.006& 0.014$\pm$ 0.007& -& 0.025$\pm$ 0.033\\
         156&00013265016 & 0.016$\pm$ 0.013& -& 0.003$\pm$ 0.003& 0.012$\pm$ 0.005& 0.028$\pm$ 0.010& 0.019$\pm$ 0.019& -\\
         172&00013265017 & 0.007$\pm$ 0.012& 0.008$\pm$ 0.001& -& -& -& -& -\\
         200&00013265019 & 0.002$\pm$ 0.009& 0.005$\pm$ 0.001& -& -& -& -& -\\
         214&00013265020 & 0.022$\pm$ 0.022& 0.004$\pm$ 0.003& -& -& -& -& -\\
         228&00013265021 & 0.001$\pm$ 0.004& 0.005$\pm$ 0.001& -& -& -& -& -\\
\hline
\end{tabular}
\tablecomments{Time is in days since 2020 February 5, the date on which the transient was first detected by \swiftxrt.}
\end{center}
\end{table*}

\begin{table}
\centering
\caption{ZTF fluxes.}
\label{table_ztf}
\begin{center}
\begin{tabular}{c c c c }
\hline
Time & g & r & i\\
(Days) & ($mJy$) & ($mJy$) & ($mJy$) \\
\hline
         -53& 0.060$\pm$ 0.012& -& 0.046$\pm$ 0.006\\
         -50& -& 0.053$\pm$ 0.010& -\\
         -48& -& -& 0.050$\pm$ 0.004\\
         -38& 0.033$\pm$ 0.002& 0.035$\pm$ 0.004& -\\
         -35& -& -& 0.039$\pm$ 0.006\\
         -30& 0.036$\pm$ 0.007& -& 0.045$\pm$ 0.008\\
         -29& 0.031$\pm$ 0.007& 0.023$\pm$ 0.004& -\\
         -28& -& 0.022$\pm$ 0.003& -\\
         -24& -& -& 0.024$\pm$ 0.008\\
         -23& 0.024$\pm$ 0.007& -& -\\
         -22& 0.023$\pm$ 0.007& -& 0.027$\pm$ 0.004\\
         -21& 0.023$\pm$ 0.005& 0.018$\pm$ 0.004& 0.025$\pm$ 0.003\\
         -20& -& -& 0.026$\pm$ 0.003\\
         -18& 0.019$\pm$ 0.003& 0.014$\pm$ 0.004& -\\
         -14& 0.013$\pm$ 0.003& -& -\\
         -13& 0.014$\pm$ 0.003& 0.014$\pm$ 0.004& 0.017$\pm$ 0.005\\
         -12& -& 0.018$\pm$ 0.003& 0.028$\pm$ 0.004\\
          -9& -& -& 0.020$\pm$ 0.003\\
          -8& 0.012$\pm$ 0.002& 0.009$\pm$ 0.002& 0.018$\pm$ 0.003\\
          -7& 0.010$\pm$ 0.003& 0.009$\pm$ 0.003& 0.017$\pm$ 0.004\\
          -6& 0.015$\pm$ 0.003& 0.015$\pm$ 0.004& 0.017$\pm$ 0.005\\
          -5& 0.010$\pm$ 0.002& -& -\\
          -4& -& 0.008$\pm$ 0.003& 0.016$\pm$ 0.003\\
          -2& -& -& 0.023$\pm$ 0.005\\
          -1& 0.021$\pm$ 0.005& 0.022$\pm$ 0.004& 0.016$\pm$ 0.005\\
           0& 0.016$\pm$ 0.005& -& -\\
           1& 0.016$\pm$ 0.005& 0.010$\pm$ 0.003& 0.014$\pm$ 0.004\\
           2& 0.026$\pm$ 0.006& 0.012$\pm$ 0.004& -\\
           3& -& 0.015$\pm$ 0.005& 0.022$\pm$ 0.006\\
           7& -& -& 0.014$\pm$ 0.003\\
           8& -& 0.012$\pm$ 0.004& -\\
           9& -& -& 0.012$\pm$ 0.003\\
          10& -& 0.012$\pm$ 0.004& -\\
          11& -& -& 0.013$\pm$ 0.003\\
          19& -& -& 0.011$\pm$ 0.003\\
          23& -& -& 0.013$\pm$ 0.003\\
          27& -& -& 0.011$\pm$ 0.003\\
          78& 0.006$\pm$ 0.002& -& -\\
          79& -& 0.010$\pm$ 0.003& -\\
          80& 0.008$\pm$ 0.002& -& -\\
\hline
\end{tabular}
\tablecomments{Time is in days since 2020 February 5, the date on which the transient was first detected by \swiftxrt.}
\end{center}
\end{table}

We show a fit to the \swiftxrt, \swiftuvot, and ZTF lightcurves in Figure \ref{fig_fit_swift_ltcrv}, and Figure \ref{con_fit_swift_ltcrv} shows the \chisq\ contours of $t_{\rm0}$ vs. $n$. We find that in X-rays, to 1$\sigma$, the power-law index is consistent with $-1.1>n>-1.9$, with a best fit of $n=-1.7$. In the UV and optical bands, the data are not as constraining and are consistent with the X-ray with e.g. $-1.1>n>-2.2$ for {\it UVW2}. There are indications that the transient in the {\it UVW2} band started prior to the X-rays, where $-45<t_{\rm0}<-5$ for X-rays and $-100<t_{\rm0}<-20$ for {\it UVW2}, although their 1$\sigma$ confidence intervals are overlapping. While the $t_{\rm0}$ constraints from the X-rays are consistent with the \erosita\ detection at $t=-40$ days, the \erosita\ flux measurement is clearly not consistent with the fit to the \swift\ lightcurve, as seen in Figure \ref{fig_fit_swift_ltcrv}.

For the ZTF data, we find that the power-law index is consistent with $-2.0<n<-1.2$, and therefore with the X-ray and UV constraints, but $-100<t_{\rm0}<-70$, which is consistent with the UV constraints, but not the X-ray ones. The transient was first detected by ZTF at $t=-53$, but could have started as early as $t=-123$ due to an observing gap. The average best fit of $t_{\rm0}$ in the $g$ and $r$ bands is $-70$ days. If the start time of the optical transient was $t=-53$, then it would be marginally consistent with the X-ray constraints for $t_{\rm0}$, but in conclusion, we do not have good constraints on when the transient started, neither in X-ray nor in the optical/UV. We summarize the lightcurve fitting results in Table \ref{tab_fitltcrv}.

\begin{table}
\centering
\caption{Lightcurve fitting results}
\label{tab_fitltcrv}
\begin{center}
\begin{tabular}{l l l l l}
\hline
Band  & $n$ &  $t_{\rm 0}$ (days) \\
\hline
X-ray (2 keV) & $-1.7_{-0.1}^{+0.2}$ & $-30^{+10}_{-5}$ \\
UV (UVW2) & $-1.9_{-0.1}^{+0.5}$ & $-80^{+40}_{-20}$ \\
Optical (i) & $-1.6_{-0.1}^{+0.2}$ & $<-$90 \\

\hline
\end{tabular}
\tablecomments{Results from the fit of a powerlaw decline model to the X-ray, UV, and optical lightcurves of the transient in \galaxy.}
\end{center}
\end{table}

We then assume that the optical, UV and X-ray transients had the same start time. We do this by fixing $t_{\rm0}$ to $-70$ days in all our lightcurve fits which is the best constraint from ZTF. This best-fit is shown as a dashed line in Figure \ref{fig_fit_swift_ltcrv} which shows it as fitting the UVOT data well. In the X-rays, it under-predicts the XRT data between 0--50 days, with a flatter power-law index, $n=-1.5$. Interestingly, this model matches the eROSITA flux better. 

\begin{figure}
\begin{center}
\includegraphics[width=80mm]{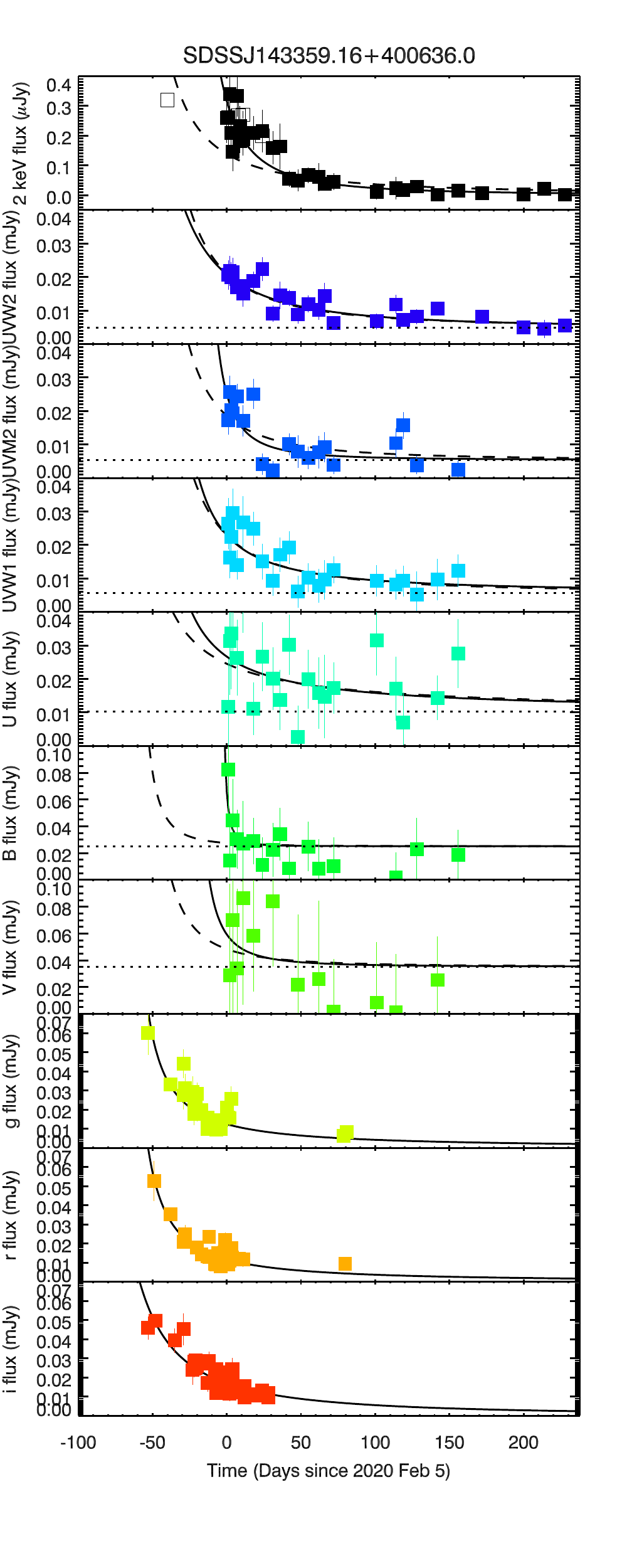}
\caption{\swiftxrt\ (2 keV), \swiftuvot\ ({\it UVW2}, {\it UVM2}, {\it UVW1}, {\it U}, {\it B}, and {\it V}), and ZTF ($g$, $r$, and $i$) lightcurves of the X-ray transient in \galaxy. Solid black lines represent fits to the data with a power-law model where all fit parameters are free to vary. Dashed black lines represent fits where the start time of the transient has been fixed to $-70$ days. Open squares in the X-ray lightcurves are the data points from \erosita, \nustar, and \chandra\ which were not used to fit the lightcurve. Black dotted lines show the quiescent flux from the galaxy in the \swiftuvot\ filters before detection of the X-ray transient.}
\label{fig_fit_swift_ltcrv}
\end{center}
\end{figure}

\begin{figure}
\begin{center}
\includegraphics[width=90mm]{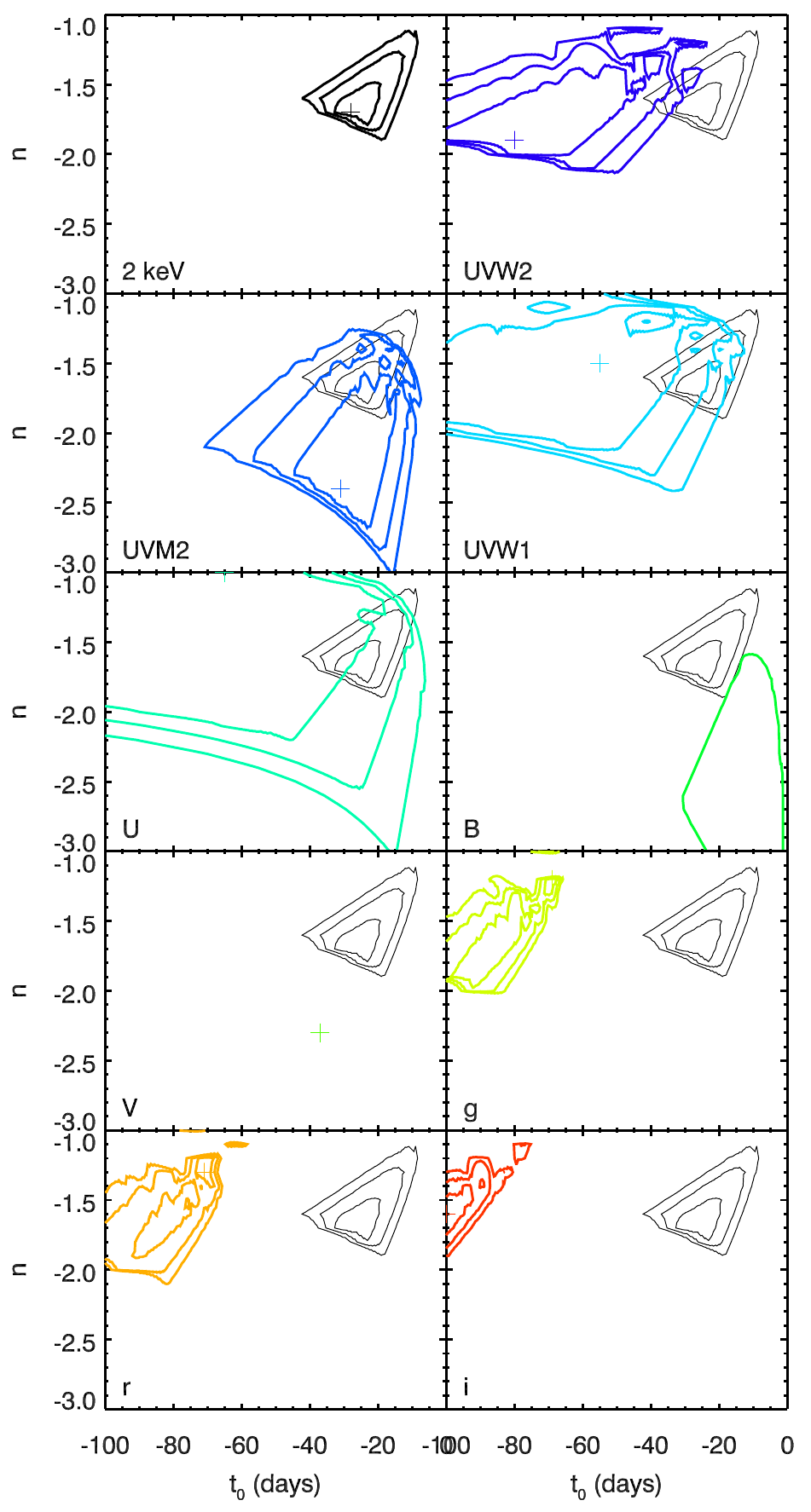}
\caption{1, 2, and 3$\sigma$ \chisq\ contours of the fits to the \swiftxrt, \swiftuvot\ and ZTF lightcurves. Crosses mark the \chisq\ minimum. The X-ray contours, plotted with black lines, are over-plotted on the optical/UV ones for comparison.}
\label{con_fit_swift_ltcrv}
\end{center}
\end{figure}

\section{SED fitting}
\label{sec_sed}

The \swiftuvot\ and ZTF data in combination with the \chandra\ and \nustar\ spectra allow us to construct a broadband SED of the source. Since the \swiftuvot\ data include emission from the host galaxy, we used the photometry inferred by the model fitting described above in Section \ref{sec_ltcrv}. This naturally accounts for the host galaxy emission underlying the source which is assumed to be constant. The photometric errors were calculated by fixing all model parameters with the exception of the normalization. The SED is shown in Figure \ref{fig_fit_sed}.

\begin{figure*}
\begin{center}
\includegraphics[width=180mm]{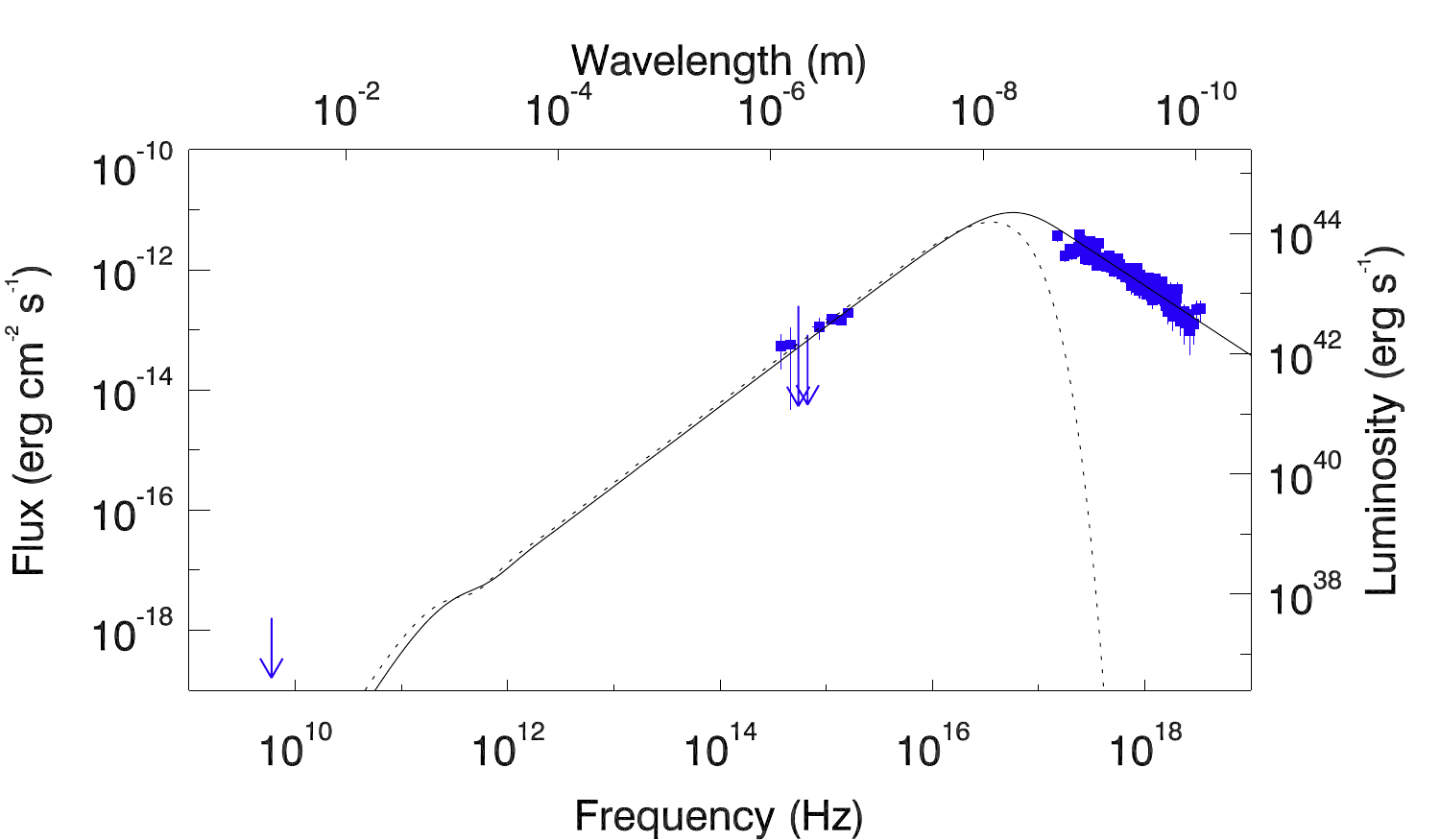}
\caption{The SED of the transient in \galaxy\ (blue data points), 8 days after it was detected by \swift, showing data from VLA, ZTF, \swiftuvot, \chandra, and \nustar. The VLA data are from 172 days after the optical--X-ray data. Upper limits are shown with downward pointing arrows. The best-fit deredened and unabsorbed disk blackbody ({\tt diskbb}, shown with a dashed line) plus powerlaw model is shown as a solid black line. }
\label{fig_fit_sed}
\end{center}
\end{figure*}

In order to fit the broadband SED, we converted the UVOT and ZTF fluxes into a PHA (pulse height amplitude) file using the tool {\sc ftflx2xsp} so that it can be loaded into {\sc xspec}. We used the time of the \nustar\ observation to calculate the UVOT photometry and take the closest ZTF data. Using {\sc xspec} and the \chisq\ statistic for spectral fitting, we find that the ZTF and \swiftuvot\ data alone can be well described by a powerlaw model, $F_{\gamma}=AE^{-\Gamma}$ where $F_{\gamma}$ is the photon flux in units of photons\,cm$^{-2}$\,s$^{-1}$\,keV$^{-1}$, $A$ is a normalization constant, $E$ is the photon energy in keV, and $\Gamma$ is the powerlaw index. For this model, we find $\Gamma=1.01^{+0.41}_{-0.56}$, where \chisq=2.42 with 6 degrees of freedom. The 0.002--0.01 keV flux is $2.2\times10^{-13}$ \ergcms, which corresponds to a luminosity of $5.1\times10^{42}$ \ergs\ at a distance of 456~Mpc, and is a factor of $\sim20$ lower than the 0.5--15 keV luminosity measured at the same time (Section \ref{sec_nustar}).

The $\Gamma=1.0$ observed in the UVOT data is much flatter than the $\Gamma=3.0$ observed in the X-ray band. For fitting the full SED, we use the \chisq\ statistic and therefore grouped the X-ray data with a minimum of 20 counts per bin. Furthermore, as with the X-ray only data, we include a Galactic absorption component, fixed at 9.8$\times10^{19}$ \cmsq\ \citep{HI4PI16}, and an intrinsic absorption component, {\tt ztbabs} accounts for absorption at the redshift of the source and is left as a free parameter. The simplest model to fit the full SED is a broken powerlaw model where the break occurs at 1 keV, which yields a good fit where \chisq=114.62 with 113 DoFs. 

We then tried fitting a more physically motivated models, specifically a standard accretion disk model, {\tt diskbb} in {\sc xspec} \citep[e.g.][]{mitsuda84,makishima86}. However this model does not produce a good fit, where \chisq/DoF=2093.68/120, fitting the ZTF and \swiftuvot\ data well, but severely under-predicting the X-ray data. 

We then introduced a scattered powerlaw in addition to the {\tt diskbb} model, using the {\tt simpl} model \citep{steiner09}. The {\tt simpl} model is an empirical convolution model of Comptonization in which a fraction of the photons in an input seed spectrum, in this case the disk black body model, is up-scattered into a power-law component. In {\sc xspec} this is written as {\tt simpl*diskbb}. This model accounts for the excess X-rays well, which significantly improves the model fit to \chisq/DoF=129.40/118. The best fit parameters of the disk model are an inner disk temperature of $T_{\rm in}=0.063^{+0.003}_{-0.007}$ keV ($7.3^{+0.3}_{-0.8}\times10^{5}$ K) and a normalization of $N=3.7^{+1.4}_{-1.0}\times10^4$. The parameters of the scattered powerlaw are $\Gamma=3.2\pm0.1$ with a scattered fraction, $f_{\rm scatt}>0.35$ (unconstrained at the upper end). The absorption intrinsic to the source is measured to be \nh$=(1.6\pm0.7)\times10^{21}$ \cmsq}. We summarize the SED fitting results in Table \ref{tab_fitsed}.

\begin{table}
\centering
\caption{SED fitting results}
\label{tab_fitsed}
\begin{center}
\begin{tabular}{l l l l l}
\hline
Parameter & Result \\
\hline
\nh & $(1.6\pm0.7)\times10^{21}$ \cmsq\ \\
$T_{\rm in}$ & $0.063^{+0.003}_{-0.007}$ keV \\
Normalization & $3.7^{+1.4}_{-1.0}\times10^4$ \\
$\Gamma$ & $3.2\pm0.1$ \\
$f_{\rm scatt}$ & $>0.35$ \\
Flux (bolometric) & ($2.3\pm0.3)\times10^{-11}$ \ergcms\ \\
Luminosity (bolometric) & $(5.7\pm0.1)\times10^{44}$ \ergs\ \\
\chisq\ & 129.40 \\
DoFs & 118 \\ 
\hline
\end{tabular}
\tablecomments{Results from the fit of a disk black body plus scattered powerlaw model to the ZTF, \swiftuvot, \nustar\ and \chandra\ data on the transient in \galaxy\ as measured 2020 February 13, 8 days after the X-ray transient was first detected by \swift.}
\end{center}
\end{table}

The normalization of the {\tt diskbb} model described above is related to the apparent inner disk radius, where $N=(R_{\rm in}/D_{\rm 10})^2 cos\theta$. $N$ is the normalization measured, $D_{\rm 10}$ is the distance to the source in units of 10 kpc, and $\theta$ the angle of the disk ($\theta= 0$ is face-on). Assuming a face-on disk and a distance of 456 Mpc to the source, the normalization of $3.7^{+1.4}_{-1.0}\times10^4$ measured corresponds to an inner disk radius of $8.3\times10^{9}$ km, or $3.8\times10^{10}$ km for an edge-on disk ($\theta=$ 87\degree). The gravitational radius of a black hole with a mass of log(\mbh/\msol)$=7.41\pm0.41$ is $3.8\times10^{10}$ km. Therefore, the implied inner disk radius is consistent with the gravitational radius of the SMBH, given the uncertainty in the mass and the unknown disk inclination.

In addition to ruling out a relativistic jet from this source from the non-detection of radio emission, models of synchrotron emission, such as {\tt srcut} and {\tt sresc} in {\sc xspec} can reproduce the ZTF and \swiftuvot\ data, but have too much curvature in the X-ray band to fit the overall SED well. A Bremsstrahlung model, such as {\tt bremss}, also does not fit the spectrum well, being too steep for the ZTF and \swiftuvot\ data and with too much curvature in the X-ray band. We therefore adopt the {\tt simpl*diskbb} as our best-fit model.

In order to calculate the bolometric luminosity of the event, we integrated the flux of the unabsorbed/deredened disk blackbody plus scattered power-law model over the 0.001--10 keV range. For the data taken at 8 days after the X-ray transient was detected by \swift\ described above, this yields $2.3\pm0.3\times10^{-11}$ \ergcms, which corresponds to a luminosity of $5.7\pm0.1\times10^{44}$ \ergs\ at a distance of 456~Mpc. Given the black hole mass of log(\mbh/\msol)$=7.41\pm0.41$ as measured from the stellar velocity dispersion, the Eddington luminosity of the SMBH is $3.1\times10^{45}$ \ergs, therefore the Eddington fraction at this time was $\sim$10\%. However, if we extrapolate the data back to when ZTF first detected the transient, when it was approximately five times more luminous in the optical bands, this implies that the Eddington fraction could have reached as high as 50\%, if not greater. 

\section{The host galaxy \galaxy}

\galaxy\ is listed in SDSS with magnitudes $u=20.76$, $g=19.23$, $r=18.56$, $i=18.21$, and $z=17.98$ \citep{alam15}, and in PanSTARRS with magnitudes $g=18.72$, $r=19.36$, $i=18.97$, $z=18.87$, and $y=18.49$ \citep{chambers16}. In the infrared, {\it WISE} measured $W1=15.67$, $W2=15.43$, $W3=12.45$, and $W4<8.86$, and in the UV {\it GALEX} measured NUV$=22.31$ \citep{bianchi11}. 

As described in Section \ref{sec_vla}, neither the transient nor the galaxy were detected in the radio, with a 3$\sigma$ upper limit of 28~$\mu$Jy on the 6 GHz flux density. The VLA Faint Images of the Radio Sky at Twenty-cm \citep[FIRST][]{becker95} survey, which covered the region with a sensitivity of 1~mJy at 1.4 GHz, also did not detect the galaxy. 

No morphological type for the galaxy is reported in the literature, however, our Tractor fitting of the ZTF images finds that its brightness profile was better fit by a deVaucoleur profile than an exponential profile, implying that it is an elliptical galaxy. 

The {\it WISE} colors of W1$-$W2=0.24 are less than the W1$-$W2$\geq0.8$ selection criterion of \cite{stern12} for AGN, meaning there was no evidence for the presence of a powerful AGN from the infrared in the galaxy prior to the X-ray transient. However, as described in Section \ref{sec_keck}, the optical line ratios revealed LINER activity in the nucleus.

\galaxy also has a companion galaxy, SDSS J143357.57+400647.3, which has an angular separation of 21\arcsec\ and has spectroscopic redshift of 0.0990 from SDSS. This angular distance corresponds to a projected separation of 38 kpc at this redshift meaning that the two galaxies are likely interacting. The companion is brighter and visually larger on the sky, implying it is the more massive of the two. 

\section{The nature of the X-ray transient in \galaxy}

The X-ray transient in \galaxy, with a peak luminosity of $\sim10^{44}$ \ergs\ and spatially coincident with the nucleus of the galaxy, is likely caused by an AGN flare or a TDE. Such events can be challenging to distinguish from each other \citep{auchettl18}. One specific example is IC 3599 which exhibited multiple X-ray flares and were interpreted both as AGN flares \citep{gruppe15} and multiple tidal disruption flares \citep{campana15}. We explore the likelihood of an AGN flare or a TDE in the following sections.

\subsection{An AGN flare in \galaxy?}

One of the distinguishing features of the X-ray transient in \galaxy\ is that the X-ray spectrum is soft, with $\Gamma\sim3$, and the spectral shape does not appear to vary with time, even as the source luminosity dropped by an order of magnitude (Figure \ref{fig_gamma}). These properties are in contrast to typical AGN properties, where the mean spectral index is $\Gamma=1.8$ \cite[e.g.][]{ricci17b}, i.e. harder than observed for this transient. Furthermore, luminous AGN usually show spectral evolution with a softer when brighter behaviour \citep[e.g.][]{sobolewska09,auchettl18}, not seen for this source. 

\begin{figure}
\begin{center}
\includegraphics[width=90mm]{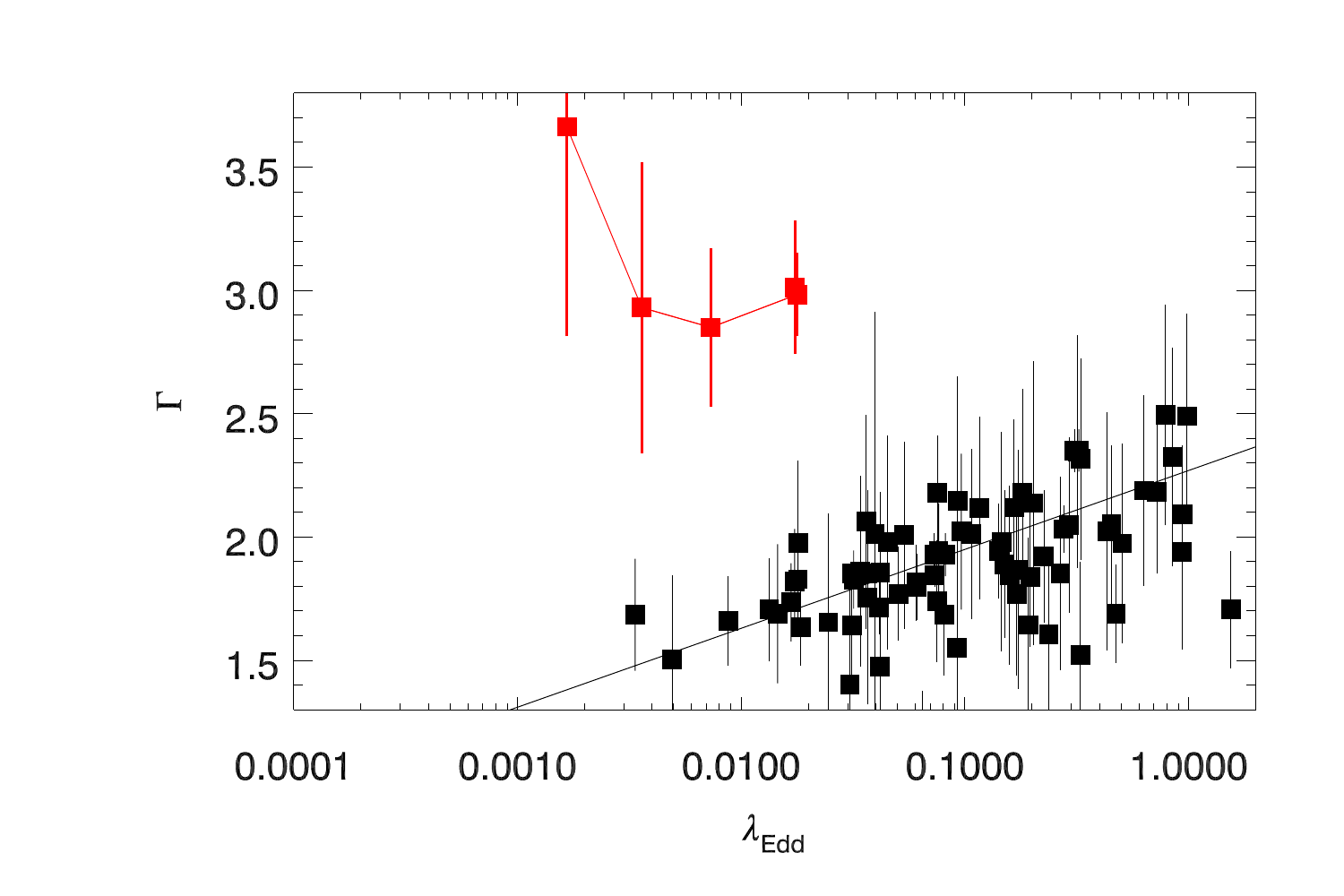}
\caption{The X-ray power-law index, $\Gamma$, of the X-ray transient in \galaxy\ plotted against its Eddington ratio, \lamedd, and how it has varied over time (red data points). Data from a sample of AGN presented in \cite{brightman13} are plotted for comparison (black data points), along with the statistically significant correlation found between these quantities (black line). This shows that $\Gamma$ is not consistent with this property of AGN, being too large for its \lamedd.}
\label{fig_lam_edd}
\end{center}
\end{figure}

This softer when brighter behaviour for AGN also reveals itself in studies of the correlation between the X-ray power-law index, $\Gamma$, and the Eddington ratio, \lamedd, \citep[e.g.][]{shemmer06,shemmer08,risaliti09}, but see \cite{trakhtenbrot17}. For example, from a sample of 69 X-ray bright sources in the \chandra\ Deep Field South and COSMOS surveys, \cite{brightman13} found that $\Gamma=(0.32\pm0.05)$log$_{10}$\lamedd$+(2.27\pm0.06)$. Given the observed peak Eddington ratio of 10\% that we have calculated, $\Gamma$ is expected to be $\sim1.8$, much lower than the value of 3 observed. We illustrate this in Figure \ref{fig_lam_edd} which shows the variation of $\Gamma$ with \lamedd\ for the X-ray transient in \galaxy\ along with the AGN data from \cite{brightman13}.  While narrow-line Seyfert 1 galaxies show steeper spectral slopes than other broad-lined AGN \citep{brandt97}, similar to our source, they also show high Eddington ratios \citep[e.g.][]{pounds95} unlike our source.

\begin{figure}
\begin{center}
\includegraphics[width=90mm]{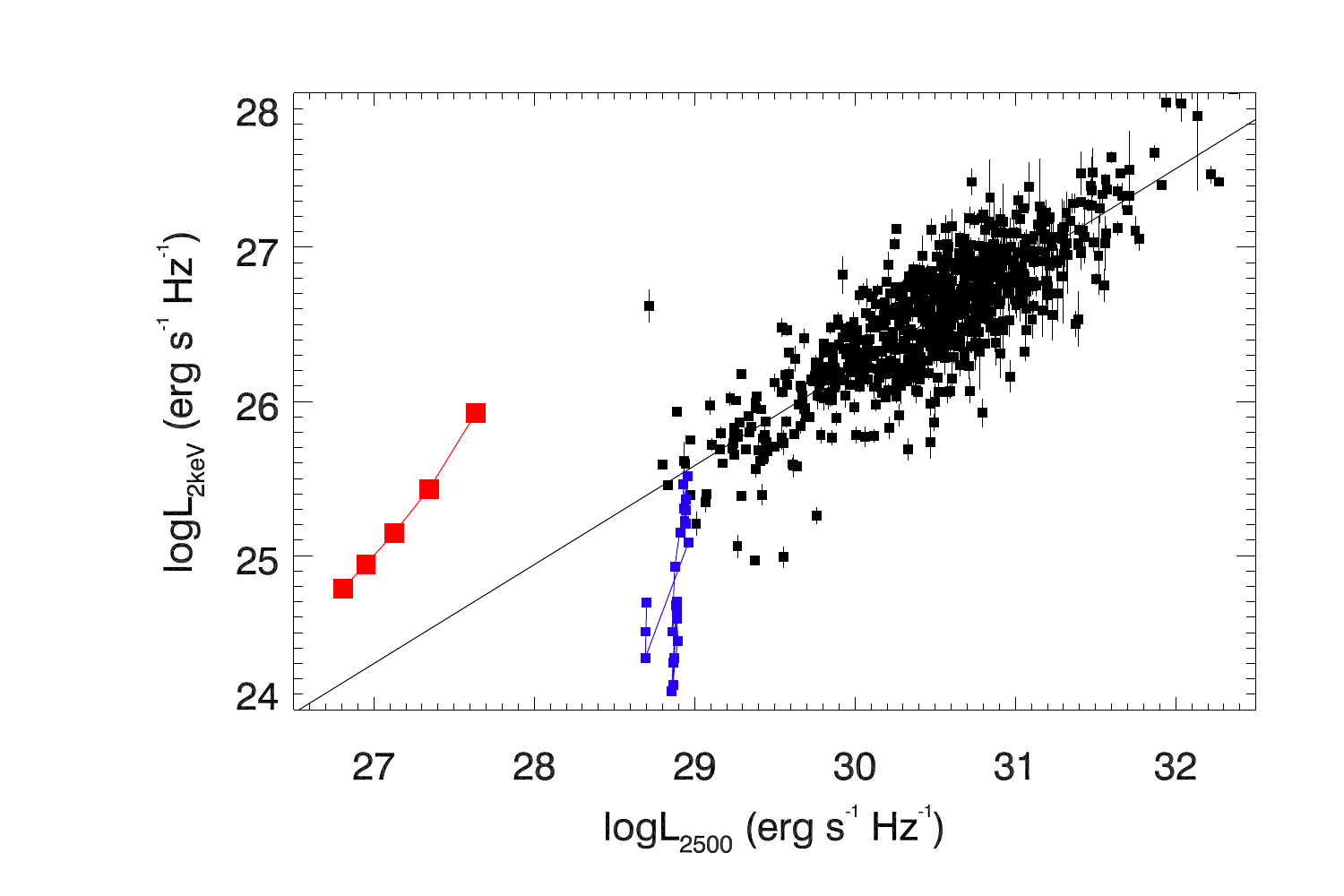}
\caption{The 2 keV luminosity of the X-ray transient in \galaxy, plotted against its luminosity at 2500\AA\ as a function of time (red data points representing data taken on 2020 Feb 5 and every 30 days after that). Data from 743 quasars selected from SDSS and 3XMM presented in \cite{lusso16} are plotted for comparison (black data points), along with the statistically significant correlation they found between these quantities (black line). Also shown are data from a flare from the AGN Mrk 335 (blue points). These show that the X-ray luminosity of the transient is not consistent with the X-ray--UV properties of quasars, being too large for its UV luminosity.}
\label{fig_luv_lx}
\end{center}
\end{figure}

Furthermore, for AGN the bright quasar-like X-ray emission should be accompanied by bright UV emission, as predicted by the tight relationship between the X-ray and UV luminosities of quasars \cite[e.g.][]{steffen06,lusso10,lusso16}. Studies of this relationship usually parameterize these quantities by the monochromatic flux densities at 2 keV and 2500 \AA. We use our fits to the lightcurve in Section \ref{sec_ltcrv} to calculate these quantities as a function of time and plot them on Figure \ref{fig_luv_lx}. Also plotted are data from 743 quasars selected from SDSS and 3XMM \citep{lusso16}, along with the relation log$L_{\rm 2 keV}=0.642L_{\rm 2500}+6.965$ derived from them.

At the observed peak of the transient, the rest frame monochromatic flux at 2 keV was 3.4$\times10^{-30}$ \ergcmshz, corresponding to a luminosity of 8.4$\times10^{25}$ \ergshz\ at $z=0.099$, whereas the flux density at 2500\AA\ as determined from our SED fit is 1.7$\times10^{-28}$ \ergcmshz, corresponding to a luminosity of 4.3$\times10^{27}$ \ergshz\ at this redshift. Given the relation log$L_{\rm 2 keV}=0.642L_{\rm 2500}+6.965$ from 743 quasars selected from SDSS and 3XMM \citep{lusso16}, the expected 2 keV luminosity of the X-ray transient in \galaxy\ given the measured 2500\AA\ one is 5.1$\times10^{24}$ \ergshz\ which an order or magnitude less luminous than measured, indicating that the X-ray transient in \galaxy\ does not exhibit the UV--X-ray properties of AGN.

However, the data from \cite{lusso16} are from single epochs observations of mostly steady-state AGN which may not capture the properties of a flaring AGN which may be more appropriate. \cite{auchettl18} conducted a comparison between a sample of X-ray TDEs and a sample of flaring AGN. The flaring AGN with most in common to \galaxy\ is Mrk 335, a narrow-line Seyfert galaxy at $z=0.025$, whose flaring activity was revealed through long-term \swift\ observations \citep[e.g.][]{gallo18}. In order to compare the X-ray to UV properties of the transient in \galaxy\ to an AGN flare, we take the \swift\ data presented in \cite{gallo18}, and plot them on Figure \ref{fig_luv_lx}. Here we have converted the XRT count rates to the 2 keV monochromatic flux density by assuming a power-law spectrum with $\Gamma=2$, and we have used the UVW1 photometry to calculate the 2500\AA\ monochromatic fluxes. The range in X-ray luminosity of the flare from Mrk 335 is comparable to that observed from \galaxy, however the UV luminosity of the flare from Mrk 335 is $\sim2$ orders of magnitude higher. This indicates that the X-ray transient in \galaxy\ does not exhibit the UV--X-ray properties of this flaring AGN.

The presence of narrow emission lines in the optical spectrum with flux ratios common to LINER galaxies suggests that a low accretion rate AGN was present in this galaxy, at least $10^{4}$ years prior to the transient this is how long it would have taken to illuminate the narrow line region located on kpc-scales from the SMBH. The galaxy would also not be selected as an AGN with its {\it WISE} colors of W1$-$W2=0.24, which is less than the W1$-$W2$\geq0.8$ criterion of \cite{assef13}. We also checked for historical AGN variability in the W1 and W2 bands by building a neoWISE \citep{mainzer11} light curve between 2014 January 8 and 2019 June18 and found no evidence of prior variability. Furthermore, the AGN luminosity inferred from the \oiii\ flux is lower than expected from the current X-ray luminosity. Therefore while a low-luminosity AGN may have existed before the onset of this new activity, it is difficult to reconcile the X-ray and UV properties of this transient with the properties of the general AGN population, or indeed an AGN flare.

\subsection{A TDE in \galaxy?}

The alternative solution is that this transient was a TDE. \cite{auchettl17} presented a comprehensive analysis of the X-ray emission from TDEs, and in \cite{auchettl18} they conducted a comparison between the X-ray properties of X-ray TDEs to flaring AGN. \cite{auchettl17} stipulated several criteria for identifying an X-ray transient as a TDE. The ones which \galaxy\ satisfies are that the X-ray light curve has a well defined shape and observable trend with several observations prior to the flare; the general shape of the X-ray light curve decay is monotonically declining; the maximum luminosity detected from the event is at least two orders of magnitude larger than the X-ray upper limit immediately preceding the discovery of the flare; over the full time range of X-ray data available for the source of interest, the candidate TDE shows evidence of X-ray emission from only the flare, while no other recurrent X-ray activity is detected; the X-ray flare is coincident with the nucleus of the host galaxy. 

One further criterion states that the X-ray light curve shows a rapid increase in X-ray luminosity, which then declines on time-scales of months to years. While the decline on time-scales of months was observed, the rise of the X-ray transient in \galaxy\ was not. \erosita\ detected the transient 40 days prior to \swift, but prior to that, the nearest X-ray observation to that was 4 years earlier, also by \swift. The \erosita\ measurement is also not consistent with the $t^{-5/3}$ as measured by \swift, but it is possible that this was part of the rise, and that the source peaked and declined before \swift\ detected it, or the lightcurve initially exhibited a plateau. This was seen in ASASSN-14li, where the X-ray lightcurve was constant for the first $\sim$100 days, after which is followed the $t^{-5/3}$ decline \citep{vanvelzen16,holoien16,brown17}.

Furthermore, \cite{auchettl17} stipulate that based on its optical spectrum or other means, one finds no evidence of AGN activity arising from its host galaxy. We find LINER-like line ratios in the optical spectrum of \galaxy, indicating low-level AGN activity prior to the event, so \galaxy\ does not strictly satisfy this criterion. However, we note that several other TDEs have shown indications of prior AGN activity, including ASASSN-14li, as determined from a radio detection and a narrow \oiii\ line \citep{vanvelzen16}, and those shown in Figure \ref{fig_bpt}. 

\begin{figure}
\begin{center}
\includegraphics[width=90mm]{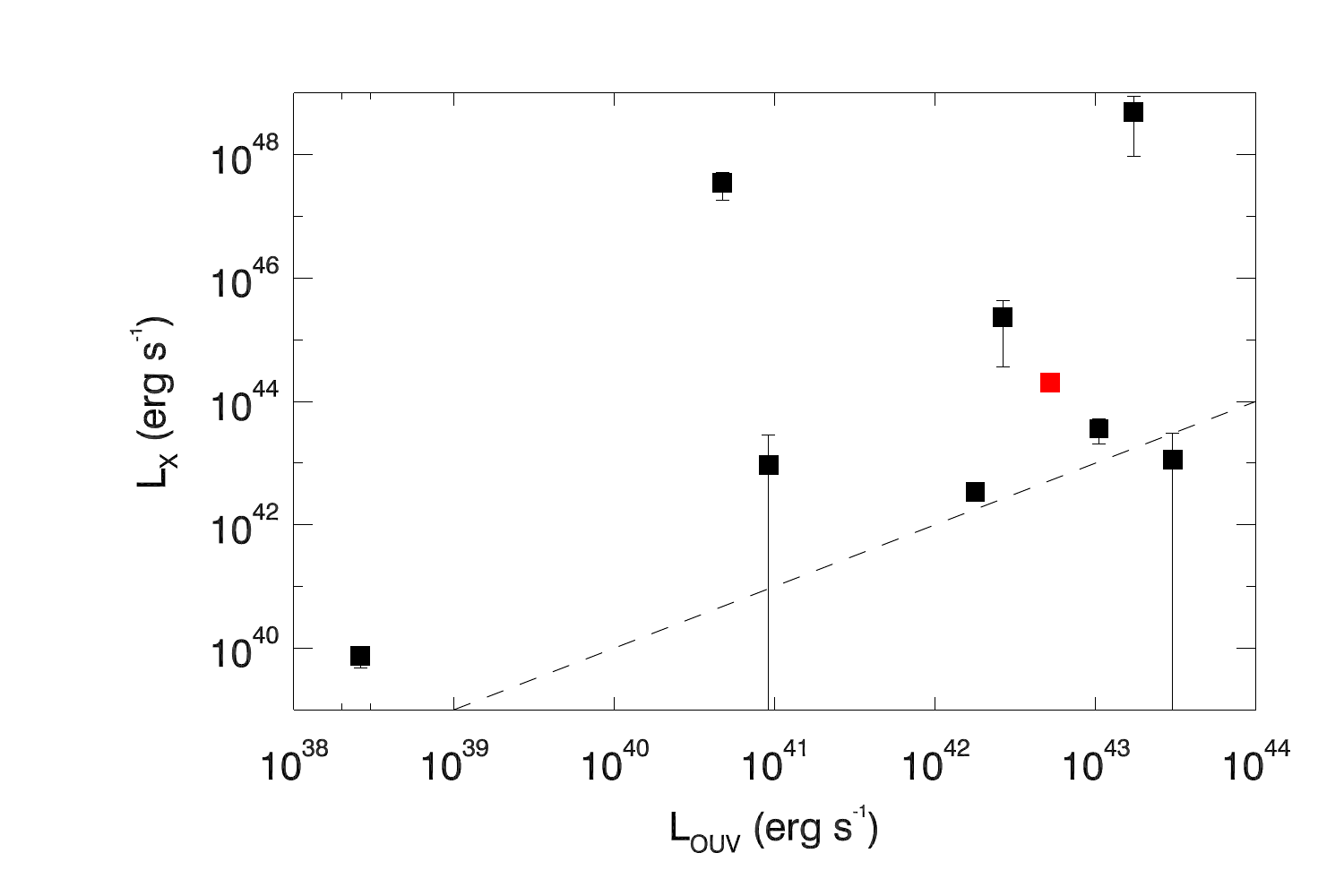}
\caption{The X-ray luminosity of the X-ray transient in \galaxy, plotted against its optical/UV luminosity (red data point). Data from a sample of X-ray TDEs presented in \cite{auchettl17} is plotted for comparison (black data points). The dashed black line marks where the two quantities are equal. The X-ray luminosity with respect to the optical/UV luminosity for \galaxy\ is consistent with other X-ray TDEs.}
\label{fig_auchettl17}
\end{center}
\end{figure}

Finally, we compare the optical/UV and X-ray luminosities of the X-ray transient in \galaxy\ to those presented for the X-ray TDEs in \cite{auchettl17} in Figure \ref{fig_auchettl17}. This shows that the X-ray luminosity with respect to the optical/UV luminosity for \galaxy\ is consistent with other X-ray TDEs, albeit that these events present more diverse properties than AGN. 

In their comparison between the X-ray properties of X-ray TDEs to flaring AGN, \cite{auchettl18} noted the lack of X-ray spectral evolution in TDEs, whereas AGN often show significant spectral evolution, as we showed in the previous section. We therefore find that since the source satisfies most of the criteria for classifying X-ray TDEs set out by \cite{auchettl17}, and that the X-ray and UV properties of the X-ray transient in \galaxy\ are more comparable to known TDEs than AGN, we conclude that the transient likely is powered by a TDE.

\section{The X-ray transient in \galaxy\ in the context of TDEs}

Of the 13 transients classified as X-ray TDEs or likely X-ray TDEs from the sample of \cite{auchettl17}, most (10) were first detected in the X-ray band, either from \xmm\ slews, serendipitously in \chandra\ or \xmm\ pointed observations, or from hard X-ray monitors such as \swift/BAT. The other three were detected in optical surveys. Therefore \galaxy\ adds to the number of TDEs first detected in the X-rays. 

In comparison to these other TDEs, we find that SDSS J1201+30 is the event which shows most similarity to \galaxy\ in terms of its X-ray and optical/UV luminosities. It was also powered by a black hole of similar mass, \citep[$10^{7.2}$ \msol,][]{wevers19}. SDSS J1201+30 was first detected by \xmm\ during a slew with \lx$\sim3\times10^{44}$ \ergs, which was 56 times brighter than a previous {\it ROSAT} upper limit and decayed with a with $t^{-5/3}$ profile \citep{saxton12}. A power-law fit to the X-ray spectrum of the source yielded $\Gamma=3.38\pm0.04$. The optical/UV emission from this source was also weak, with 0.002--0.1 keV luminosity of $2.64\pm0.31\times10^{42}$ \ergs\ \citep{auchettl17}. The source also did not present broad or coronal optical lines. The X-ray spectrum could be reproduced with a Bremsstrahlung or double-power-law model. These characteristics are similar to \galaxy. 

One property of SDSS J1201+30 that we do not see in \galaxy\ is variability on timescales of days in addition to the monotonic flux decline. SDSS J1201+30 became invisible to \swift\ between 27 and 48 days after discovery, which \cite{saxton12} suggested could be due to self-absorption by material driven from the system by radiation pressure during an early super-Eddington accretion phase. Alternatively, \cite{liu14_tde} suggested that a supermassive black hole binary lies at the heart of SDSS J1201+30, and that the dips in the lightcurve were due to disruption of the accretion flow by the secondary SMBH. \galaxy, however, does not show evidence for excess variability from the powerlaw decline. 

In terms of how the X-ray lightcurve of \galaxy\ compares with the well sampled X-ray light curves of other X-ray TDEs, \galaxy\ appears to have shown a plateau of emission before declining, similar to ASSASN-14li \citep{vanvelzen16}, while XMMSL1 J0740-85 declined monotonically without evidence for a plateau \citep{saxton17}.

Having compared the properties of the TDE in \galaxy\ to other X-ray TDEs, it is useful to compare the optical emission from the TDE in \galaxy\ to that of optically selected TDEs. For this we use the recent sample of 17 ZTF-discovered TDEs presented in \cite{vanvelzen20}. Here the authors use a simple blackbody model to fit the optical/UV data of their sample. We proceed to fit the optical/UV data described in Section \ref{sec_sed}, finding that these can be described by a blackbody with log($T$/K)$=4.3^{+0.2}_{-0.1}$, where the $g$-band luminosity is log($L_{g}$/\ergs)=41.0$\pm$0.1, and the total blackbody luminosity is log($L_{\rm bb}$/\ergs)=42.8$\pm$0.1. While the temperature is comparable to the sample of \cite{vanvelzen20}, which has the range log($T$/K)=4.1--4.6, the luminosities are much lower, where the ZTF TDEs have log($L_{g}$/\ergs)=42.8--43.6 and log($L_{\rm bb}$/\ergs)=43.2--44.7.

The black hole mass inferred from the stellar velocity dispersion of \galaxy\ is $\sim10^{7.4}$ \msol, which is around the peak of the observed distribution of black hole masses of TDEs \citep{stone16}, although for optical events this was found to be lower, $\sim10^{6}$ \msol\ \citep{wevers17}. We calculated that the Eddington fraction of the event near peak was only $\sim10$\%. This is naturally explained since the SMBH has a mass of $\sim10^{7.4}$ \msol, meaning that a very massive star would have been needed to reach Eddington luminosities. \cite{strubbe09} stated that TDEs can emit above the Eddington luminosity for a BH with \mbh$<10^{7}$ \mbh. Indeed \cite{stone16} concluded that Eddington-limited emission channels of TDEs dominate the rates.

Finally, we noted that \galaxy\ has a companion galaxy, SDSS J143357.57+400647.3, which has a projected separation of 38 kpc. This may be important since a companion galaxy that may be undergoing an interaction with the host could be relevant to the fueling of TDEs \citep{french20}.

\section{Implications and Conclusions}

Only 13 transients were classified as X-ray TDEs or likely X-ray TDEs from the sample of \cite{auchettl17}, so the number of known X-ray TDEs is still small. Therefore finding more events of this nature are important for understanding this population, even just one event as we have reported here. 

This TDE was one of a few identified where previous AGN activity in the galaxy was known, albeit at a low-level. Other TDEs with known AGN activity prior to the flare include ASSASN-14li \citep{vanvelzen16}, where archival radio data and narrow \oiii\ emission showed a low-luminosity AGN existed prior to the event. As can be seen in Fig \ref{fig_bpt}, several other TDE hosts showed similar evidence for prior AGN activity from their narrow line ratios, including CNSS J0019+00 \citep[Sy2,][]{anderson19}. Furthermore, \cite{ricci20} postulated that a TDE caused the changing-look behaviour of the AGN 1ES 1927+654, and \cite{merloni15} suggested that TDEs may be drivers of these changing-look events.

While we used ZTF data to determine the optical evolution of this TDE, this event was not identified as a TDE by wide field optical surveys such as ZTF or ASAS-SN, possibly due to its low optical luminosity. We note, however, that ZTF was not observing the field of \galaxy\ when the optical luminosity was at its peak, which may be the reason it was missed. This TDE was also not classified as a TDE from its optical spectrum. Taken together, this suggests many more events like it are being missed, and ultimately only wide field UV or X-ray surveys will catch events like these. \erosita\ is currently conducting an all-sky survey in the 0.2--10 keV band and will likely identify a large number of them \citep{merloni12}.

In conclusion, we have reported on an X-ray transient, observed to peak at a 0.3--10 keV luminosity of $10^{44}$ \ergs, originating in the nucleus of the galaxy \galaxy\ at $z=0.099$. The X-ray transient was also accompanied by a less powerful optical/UV transient. A soft X-ray spectrum with $\Gamma=3$ and the low UV/X-ray ratio disfavor an AGN flare scenario. The source was observed to decline monotonically in all bands, consistent with a $t^{-5/3}$ profile favoring a TDE scenario. Since this event was not identified as a TDE by wide-field optical surveys, or by optical spectroscopy, we are lead to the conclusion that a significant fraction of X-ray TDEs may be going unnoticed.

\facilities{Swift (XRT, UVOT), NuSTAR, CXO, Keck:I (LRIS), PO:1.2m PO:1.5m, VLA} 

\software{{\tt CASA} \citep{mcmullin07}, {\tt CIAO} \citep{fruscione06}, {\tt lpipe} \citep{perley19}, {\tt The Tractor} \citep{lang16}, {\tt XSPEC} \citep{arnaud96}}

\acknowledgements{
The majority of this research and manuscript preparation took place during the COVID-19 global pandemic. The authors would like to thank all those who risked their lives as essential workers in order for us to safely continue our work from home. 

We wish to thank the \swift\ PI, Brad Cenko, for approving the target of opportunity requests we made to observe \galaxy, as well as the rest of the \swift\ team for carrying the observations out. We also acknowledge the use of public data from the \swift\ data archive.

We also wish to thank the \nustar\ PI, Fiona Harrison, for approving the DDT request we made to observe \galaxy, as well as the \nustar\ SOC for carrying out the observation. This work was also supported under NASA Contract No. NNG08FD60C. \nustar\ is a project led by the California Institute of Technology, managed by the Jet Propulsion Laboratory, and funded by the National Aeronautics and Space Administration. This research has made use of the NuSTAR Data Analysis Software (NuSTARDAS) jointly developed by the ASI Science Data Center (ASDC, Italy) and the California Institute of Technology (USA).

ZTF is supported by the National Science Foundation under Grant No. AST-1440341 and a collaboration including Caltech, IPAC, the Weizmann Institute for Science, the Oskar Klein Center at Stockholm University, the University of Maryland, the University of Washington, Deutsches Elektronen-Synchrotron and Humboldt University, Los Alamos National Laboratories, the TANGO Consortium of Taiwan, the University of Wisconsin at Milwaukee, and Lawrence Berkeley National Laboratories. Operations are conducted by COO, IPAC, and UW. 

This paper is based on observations obtained with the Samuel Oschin Telescope 48-inch and the 60-inch Telescope at the Palomar Observatory as part of the Zwicky Transient Facility project. ZTF is supported by the National Science Foundation under Grant No. AST-1440341 and a collaboration including Caltech, IPAC, the Weizmann Institute for Science, the Oskar Klein Center at Stockholm University, the University of Maryland, the University of Washington, Deutsches Elektronen-Synchrotron and Humboldt University, Los Alamos National Laboratories, the TANGO Consortium of Taiwan, the University of Wisconsin at Milwaukee, and Lawrence Berkeley National Laboratories. Operations are conducted by COO, IPAC, and UW.  SED Machine is based upon work supported by the National Science Foundation under Grant No. 1106171 .

The ZTF forced-photometry service was funded under the Heising-Simons Foundation grant \#12540303 (PI: Graham).}

\bibliography{manuscript.bbl}

\begin{thebibliography}{}
\expandafter\ifx\csname natexlab\endcsname\relax\def\natexlab#1{#1}\fi

\bibitem[{{Alam} {et~al.}(2015){Alam}, {Albareti}, {Allende Prieto}, {Anders},
  {Anderson}, {Anderton}, {Andrews}, {Armengaud}, {Aubourg}, {Bailey}, {Basu},
  {Bautista}, {Beaton}, {Beers}, {Bender}, {Berlind}, {Beutler}, {Bhardwaj},
  {Bird}, {Bizyaev}, {Blake}, {Blanton}, {Blomqvist}, {Bochanski}, {Bolton},
  {Bovy}, {Shelden Bradley}, {Brandt}, {Brauer}, {Brinkmann}, {Brown},
  {Brownstein}, {Burden}, {Burtin}, {Busca}, {Cai}, {Capozzi}, {Carnero
  Rosell}, {Carr}, {Carrera}, {Chambers}, {Chaplin}, {Chen}, {Chiappini},
  {Chojnowski}, {Chuang}, {Clerc}, {Comparat}, {Covey}, {Croft}, {Cuesta},
  {Cunha}, {da Costa}, {Da Rio}, {Davenport}, {Dawson}, {De Lee}, {Delubac},
  {Deshpande}, {Dhital}, {Dutra-Ferreira}, {Dwelly}, {Ealet}, {Ebelke},
  {Edmondson}, {Eisenstein}, {Ellsworth}, {Elsworth}, {Epstein}, {Eracleous},
  {Escoffier}, {Esposito}, {Evans}, {Fan}, {Fern{\'a}ndez-Alvar}, {Feuillet},
  {Filiz Ak}, {Finley}, {Finoguenov}, {Flaherty}, {Fleming}, {Font-Ribera},
  {Foster}, {Frinchaboy}, {Galbraith-Frew}, {Garc{\'\i}a},
  {Garc{\'\i}a-Hern{\'a}ndez}, {Garc{\'\i}a P{\'e}rez}, {Gaulme}, {Ge},
  {G{\'e}nova-Santos}, {Georgakakis}, {Ghezzi}, {Gillespie}, {Girardi},
  {Goddard}, {Gontcho}, {Gonz{\'a}lez Hern{\'a}ndez}, {Grebel}, {Green},
  {Grieb}, {Grieves}, {Gunn}, {Guo}, {Harding}, {Hasselquist}, {Hawley},
  {Hayden}, {Hearty}, {Hekker}, {Ho}, {Hogg}, {Holley-Bockelmann}, {Holtzman},
  {Honscheid}, {Huber}, {Huehnerhoff}, {Ivans}, {Jiang}, {Johnson},
  {Kinemuchi}, {Kirkby}, {Kitaura}, {Klaene}, {Knapp}, {Kneib}, {Koenig},
  {Lam}, {Lan}, {Lang}, {Laurent}, {Le Goff}, {Leauthaud}, {Lee}, {Lee},
  {Licquia}, {Liu}, {Long}, {L{\'o}pez-Corredoira}, {Lorenzo-Oliveira},
  {Lucatello}, {Lundgren}, {Lupton}, {Mack}, {Mahadevan}, {Maia}, {Majewski},
  {Malanushenko}, {Malanushenko}, {Manchado}, {Manera}, {Mao}, {Maraston},
  {Marchwinski}, {Margala}, {Martell}, {Martig}, {Masters}, {Mathur},
  {McBride}, {McGehee}, {McGreer}, {McMahon}, {M{\'e}nard}, {Menzel},
  {Merloni}, {M{\'e}sz{\'a}ros}, {Miller}, {Miralda-Escud{\'e}}, {Miyatake},
  {Montero-Dorta}, {More}, {Morganson}, {Morice-Atkinson}, {Morrison},
  {Mosser}, {Muna}, {Myers}, {Nand ra}, {Newman}, {Neyrinck}, {Nguyen},
  {Nichol}, {Nidever}, {Noterdaeme}, {Nuza}, {O'Connell}, {O'Connell},
  {O'Connell}, {Ogando}, {Olmstead}, {Oravetz}, {Oravetz}, {Osumi}, {Owen},
  {Padgett}, {Padmanabhan}, {Paegert}, {Palanque-Delabrouille}, {Pan},
  {Parejko}, {P{\^a}ris}, {Park}, {Pattarakijwanich}, {Pellejero-Ibanez},
  {Pepper}, {Percival}, {P{\'e}rez-Fournon}, {{\'P}rez-Ra`fols}, {Petitjean},
  {Pieri}, {Pinsonneault}, {Porto de Mello}, {Prada}, {Prakash},
  {Price-Whelan}, {Protopapas}, {Raddick}, {Rahman}, {Reid}, {Rich}, {Rix},
  {Robin}, {Rockosi}, {Rodrigues}, {Rodr{\'\i}guez-Torres}, {Roe}, {Ross},
  {Ross}, {Rossi}, {Ruan}, {Rubi{\~n}o-Mart{\'\i}n}, {Rykoff},
  {Salazar-Albornoz}, {Salvato}, {Samushia}, {S{\'a}nchez}, {Santiago},
  {Sayres}, {Schiavon}, {Schlegel}, {Schmidt}, {Schneider}, {Schultheis},
  {Schwope}, {Sc{\'o}ccola}, {Scott}, {Sellgren}, {Seo}, {Serenelli}, {Shane},
  {Shen}, {Shetrone}, {Shu}, {Silva Aguirre}, {Sivarani}, {Skrutskie},
  {Slosar}, {Smith}, {Sobreira}, {Souto}, {Stassun}, {Steinmetz}, {Stello},
  {Strauss}, {Streblyanska}, {Suzuki}, {Swanson}, {Tan}, {Tayar}, {Terrien},
  {Thakar}, {Thomas}, {Thomas}, {Thompson}, {Tinker}, {Tojeiro}, {Troup},
  {Vargas-Maga{\~n}a}, {Vazquez}, {Verde}, {Viel}, {Vogt}, {Wake}, {Wang},
  {Weaver}, {Weinberg}, {Weiner}, {White}, {Wilson}, {Wisniewski},
  {Wood-Vasey}, {Ye`che}, {York}, {Zakamska}, {Zamora}, {Zasowski}, {Zehavi},
  {Zhao}, {Zheng}, {Zhou}, {Zhou}, {Zou}, \& {Zhu}}]{alam15}
{Alam}, S., {Albareti}, F.~D., {Allende Prieto}, C., {et~al.} 2015, \apjs, 219,
  12

\bibitem[{{Alexander} {et~al.}(2020){Alexander}, {van Velzen}, {Horesh}, \&
  {Zauderer}}]{alexander2020}
{Alexander}, K.~D., {van Velzen}, S., {Horesh}, A., \& {Zauderer}, B.~A. 2020,
  \ssr, 216, 81

\bibitem[{{Anderson} {et~al.}(2019){Anderson}, {Mooley}, {Hallinan}, {Dong},
  {Phinney}, {Horesh}, {Bourke}, {Cenko}, {Frail}, {Kulkarni}, \&
  {Myers}}]{anderson19}
{Anderson}, M.~M., {Mooley}, K.~P., {Hallinan}, G., {et~al.} 2019, arXiv
  e-prints, arXiv:1910.11912

\bibitem[{{Arnaud}(1996)}]{arnaud96}
{Arnaud}, K.~A. 1996, in Astronomical Society of the Pacific Conference Series,
  Vol. 101, Astronomical Data Analysis Software and Systems V, ed. G.~H.
  {Jacoby} \& J.~{Barnes}, 17--+

\bibitem[{{Assef} {et~al.}(2013){Assef}, {Stern}, {Kochanek}, {Blain},
  {Brodwin}, {Brown}, {Donoso}, {Eisenhardt}, {Jannuzi}, {Jarrett}, {Stanford},
  {Tsai}, {Wu}, \& {Yan}}]{assef13}
{Assef}, R.~J., {Stern}, D., {Kochanek}, C.~S., {et~al.} 2013, \apj, 772, 26

\bibitem[{{Auchettl} {et~al.}(2017){Auchettl}, {Guillochon}, \&
  {Ramirez-Ruiz}}]{auchettl17}
{Auchettl}, K., {Guillochon}, J., \& {Ramirez-Ruiz}, E. 2017, \apj, 838, 149

\bibitem[{{Auchettl} {et~al.}(2018){Auchettl}, {Ramirez-Ruiz}, \&
  {Guillochon}}]{auchettl18}
{Auchettl}, K., {Ramirez-Ruiz}, E., \& {Guillochon}, J. 2018, \apj, 852, 37

\bibitem[{{Becker} {et~al.}(1995){Becker}, {White}, \& {Helfand}}]{becker95}
{Becker}, R.~H., {White}, R.~L., \& {Helfand}, D.~J. 1995, \apj, 450, 559

\bibitem[{{Bellm} {et~al.}(2019){Bellm}, {Kulkarni}, {Graham}, {Dekany},
  {Smith}, {Riddle}, {Masci}, {Helou}, {Prince}, {Adams}, {Barbarino},
  {Barlow}, {Bauer}, {Beck}, {Belicki}, {Biswas}, {Blagorodnova}, {Bodewits},
  {Bolin}, {Brinnel}, {Brooke}, {Bue}, {Bulla}, {Burruss}, {Cenko}, {Chang},
  {Connolly}, {Coughlin}, {Cromer}, {Cunningham}, {De}, {Delacroix}, {Desai},
  {Duev}, {Eadie}, {Farnham}, {Feeney}, {Feindt}, {Flynn}, {Franckowiak},
  {Frederick}, {Fremling}, {Gal-Yam}, {Gezari}, {Giomi}, {Goldstein},
  {Golkhou}, {Goobar}, {Groom}, {Hacopians}, {Hale}, {Henning}, {Ho}, {Hover},
  {Howell}, {Hung}, {Huppenkothen}, {Imel}, {Ip}, {Ivezi{\'c}}, {Jackson},
  {Jones}, {Juric}, {Kasliwal}, {Kaspi}, {Kaye}, {Kelley}, {Kowalski},
  {Kramer}, {Kupfer}, {Landry}, {Laher}, {Lee}, {Lin}, {Lin}, {Lunnan},
  {Giomi}, {Mahabal}, {Mao}, {Miller}, {Monkewitz}, {Murphy}, {Ngeow},
  {Nordin}, {Nugent}, {Ofek}, {Patterson}, {Penprase}, {Porter}, {Rauch},
  {Rebbapragada}, {Reiley}, {Rigault}, {Rodriguez}, {van Roestel}, {Rusholme},
  {van Santen}, {Schulze}, {Shupe}, {Singer}, {Soumagnac}, {Stein}, {Surace},
  {Sollerman}, {Szkody}, {Taddia}, {Terek}, {Van Sistine}, {van Velzen},
  {Vestrand}, {Walters}, {Ward}, {Ye}, {Yu}, {Yan}, \& {Zolkower}}]{bellm19}
{Bellm}, E.~C., {Kulkarni}, S.~R., {Graham}, M.~J., {et~al.} 2019, \pasp, 131,
  018002

\bibitem[{{Berney} {et~al.}(2015){Berney}, {Koss}, {Trakhtenbrot}, {Ricci},
  {Lamperti}, {Schawinski}, {Balokovi{\'c}}, {Crenshaw}, {Fischer}, {Gehrels},
  {Harrison}, {Hashimoto}, {Ichikawa}, {Mushotzky}, {Oh}, {Stern}, {Treister},
  {Ueda}, {Veilleux}, \& {Winter}}]{berney15}
{Berney}, S., {Koss}, M., {Trakhtenbrot}, B., {et~al.} 2015, \mnras, 454, 3622

\bibitem[{{Bianchi} {et~al.}(2011){Bianchi}, {Herald}, {Efremova}, {Girardi},
  {Zabot}, {Marigo}, {Conti}, \& {Shiao}}]{bianchi11}
{Bianchi}, L., {Herald}, J., {Efremova}, B., {et~al.} 2011, \apss, 335, 161

\bibitem[{{Boller} {et~al.}(2016){Boller}, {Freyberg}, {Tr{\"u}mper}, {Haberl},
  {Voges}, \& {Nandra}}]{boller16}
{Boller}, T., {Freyberg}, M.~J., {Tr{\"u}mper}, J., {et~al.} 2016, \aap, 588,
  A103

\bibitem[{{Brandt} {et~al.}(1997){Brandt}, {Mathur}, \& {Elvis}}]{brandt97}
{Brandt}, W.~N., {Mathur}, S., \& {Elvis}, M. 1997, \mnras, 285, L25

\bibitem[{{Brightman} {et~al.}(2013){Brightman}, {Silverman}, {Mainieri},
  {Ueda}, {Schramm}, {Matsuoka}, {Nagao}, \& {Steinhardt}}]{brightman13}
{Brightman}, M., {Silverman}, J.~D., {Mainieri}, V., {et~al.} 2013, \mnras,
  433, 2485

\bibitem[{{Brown} {et~al.}(2017){Brown}, {Holoien}, {Auchettl}, {Stanek},
  {Kochanek}, {Shappee}, {Prieto}, \& {Grupe}}]{brown17}
{Brown}, J.~S., {Holoien}, T.~W.~S., {Auchettl}, K., {et~al.} 2017, \mnras,
  466, 4904

\bibitem[{{Burrows} {et~al.}(2005){Burrows}, {Hill}, {Nousek}, {Kennea},
  {Wells}, {Osborne}, {Abbey}, {Beardmore}, {Mukerjee}, {Short}, {Chincarini},
  {Campana}, {Citterio}, {Moretti}, {Pagani}, {Tagliaferri}, {Giommi},
  {Capalbi}, {Tamburelli}, {Angelini}, {Cusumano}, {Br{\"a}uninger}, {Burkert},
  \& {Hartner}}]{burrows05}
{Burrows}, D.~N., {Hill}, J.~E., {Nousek}, J.~A., {et~al.} 2005, Space Science
  Reviews, 120, 165

\bibitem[{{Campana} {et~al.}(2015){Campana}, {Mainetti}, {Colpi}, {Lodato},
  {D'Avanzo}, {Evans}, \& {Moretti}}]{campana15}
{Campana}, S., {Mainetti}, D., {Colpi}, M., {et~al.} 2015, \aap, 581, A17

\bibitem[{{Cappellari}(2017)}]{cappellari17}
{Cappellari}, M. 2017, \mnras, 466, 798

\bibitem[{{Cappellari} \& {Emsellem}(2004)}]{cappellari04}
{Cappellari}, M., \& {Emsellem}, E. 2004, \pasp, 116, 138

\bibitem[{{Chambers} {et~al.}(2016){Chambers}, {Magnier}, {Metcalfe},
  {Flewelling}, {Huber}, {Waters}, {Denneau}, {Draper}, {Farrow}, {Finkbeiner},
  {Holmberg}, {Koppenhoefer}, {Price}, {Rest}, {Saglia}, {Schlafly}, {Smartt},
  {Sweeney}, {Wainscoat}, {Burgett}, {Chastel}, {Grav}, {Heasley}, {Hodapp},
  {Jedicke}, {Kaiser}, {Kudritzki}, {Luppino}, {Lupton}, {Monet}, {Morgan},
  {Onaka}, {Shiao}, {Stubbs}, {Tonry}, {White}, {Ba{\~n}ados}, {Bell},
  {Bender}, {Bernard}, {Boegner}, {Boffi}, {Botticella}, {Calamida},
  {Casertano}, {Chen}, {Chen}, {Cole}, {Deacon}, {Frenk}, {Fitzsimmons},
  {Gezari}, {Gibbs}, {Goessl}, {Goggia}, {Gourgue}, {Goldman}, {Grant},
  {Grebel}, {Hambly}, {Hasinger}, {Heavens}, {Heckman}, {Henderson}, {Henning},
  {Holman}, {Hopp}, {Ip}, {Isani}, {Jackson}, {Keyes}, {Koekemoer}, {Kotak},
  {Le}, {Liska}, {Long}, {Lucey}, {Liu}, {Martin}, {Masci}, {McLean}, {Mindel},
  {Misra}, {Morganson}, {Murphy}, {Obaika}, {Narayan}, {Nieto-Santisteban},
  {Norberg}, {Peacock}, {Pier}, {Postman}, {Primak}, {Rae}, {Rai}, {Riess},
  {Riffeser}, {Rix}, {R{\"o}ser}, {Russel}, {Rutz}, {Schilbach}, {Schultz},
  {Scolnic}, {Strolger}, {Szalay}, {Seitz}, {Small}, {Smith}, {Soderblom},
  {Taylor}, {Thomson}, {Taylor}, {Thakar}, {Thiel}, {Thilker}, {Unger},
  {Urata}, {Valenti}, {Wagner}, {Walder}, {Walter}, {Watters}, {Werner},
  {Wood-Vasey}, \& {Wyse}}]{chambers16}
{Chambers}, K.~C., {Magnier}, E.~A., {Metcalfe}, N., {et~al.} 2016, arXiv
  e-prints, arXiv:1612.05560

\bibitem[{{Chen} {et~al.}(2019){Chen}, {Shi}, {Dempsey}, {Law}, {Chen}, {Yan},
  {Bing}, {Rembold}, {Li}, {Yu}, {Riffel}, {Brownstein}, \& {Riffel}}]{chen19}
{Chen}, J., {Shi}, Y., {Dempsey}, R., {et~al.} 2019, \mnras, 489, 855

\bibitem[{{Evans} \& {Kochanek}(1989)}]{evans89}
{Evans}, C.~R., \& {Kochanek}, C.~S. 1989, \apjl, 346, L13

\bibitem[{{Evans} {et~al.}(2007){Evans}, {Beardmore}, {Page}, {Tyler},
  {Osborne}, {Goad}, {O'Brien}, {Vetere}, {Racusin}, {Morris}, {Burrows},
  {Capalbi}, {Perri}, {Gehrels}, \& {Romano}}]{evans07}
{Evans}, P.~A., {Beardmore}, A.~P., {Page}, K.~L., {et~al.} 2007, \aap, 469,
  379

\bibitem[{{Evans} {et~al.}(2009){Evans}, {Beardmore}, {Page}, {Osborne},
  {O'Brien}, {Willingale}, {Starling}, {Burrows}, {Godet}, {Vetere}, {Racusin},
  {Goad}, {Wiersema}, {Angelini}, {Capalbi}, {Chincarini}, {Gehrels}, {Kennea},
  {Margutti}, {Morris}, {Mountford}, {Pagani}, {Perri}, {Romano}, \&
  {Tanvir}}]{evans09}
---. 2009, \mnras, 397, 1177

\bibitem[{{French} {et~al.}(2016){French}, {Arcavi}, \& {Zabludoff}}]{french16}
{French}, K.~D., {Arcavi}, I., \& {Zabludoff}, A. 2016, \apjl, 818, L21

\bibitem[{{French} {et~al.}(2017){French}, {Arcavi}, \& {Zabludoff}}]{french17}
---. 2017, \apj, 835, 176

\bibitem[{{French} {et~al.}(2020){French}, {Wevers}, {Law-Smith}, {Graur}, \&
  {Zabludoff}}]{french20}
{French}, K.~D., {Wevers}, T., {Law-Smith}, J., {Graur}, O., \& {Zabludoff},
  A.~I. 2020, \ssr, 216, 32

\bibitem[{{Fruscione} {et~al.}(2006){Fruscione}, {McDowell}, {Allen},
  {Brickhouse}, {Burke}, {Davis}, {Durham}, {Elvis}, {Galle}, {Harris},
  {Huenemoerder}, {Houck}, {Ishibashi}, {Karovska}, {Nicastro}, {Noble},
  {Nowak}, {Primini}, {Siemiginowska}, {Smith}, \& {Wise}}]{fruscione06}
{Fruscione}, A., {McDowell}, J.~C., {Allen}, G.~E., {et~al.} 2006, in Society
  of Photo-Optical Instrumentation Engineers (SPIE) Conference Series, Vol.
  6270, Society of Photo-Optical Instrumentation Engineers (SPIE) Conference
  Series, ed. D.~R. {Silva} \& R.~E. {Doxsey}, 62701V

\bibitem[{{Gaia Collaboration} {et~al.}(2018){Gaia Collaboration}, {Brown},
  {Vallenari}, {Prusti}, {de Bruijne}, {Babusiaux}, {Bailer-Jones}, {Biermann},
  {Evans}, {Eyer}, {Jansen}, {Jordi}, {Klioner}, {Lammers}, {Lindegren},
  {Luri}, {Mignard}, {Panem}, {Pourbaix}, {Randich}, {Sartoretti}, {Siddiqui},
  {Soubiran}, {van Leeuwen}, {Walton}, {Arenou}, {Bastian}, {Cropper},
  {Drimmel}, {Katz}, {Lattanzi}, {Bakker}, {Cacciari}, {Casta{\~n}eda},
  {Chaoul}, {Cheek}, {De Angeli}, {Fabricius}, {Guerra}, {Holl}, {Masana},
  {Messineo}, {Mowlavi}, {Nienartowicz}, {Panuzzo}, {Portell}, {Riello},
  {Seabroke}, {Tanga}, {Th{\'e}venin}, {Gracia-Abril}, {Comoretto},
  {Garcia-Reinaldos}, {Teyssier}, {Altmann}, {Andrae}, {Audard},
  {Bellas-Velidis}, {Benson}, {Berthier}, {Blomme}, {Burgess}, {Busso},
  {Carry}, {Cellino}, {Clementini}, {Clotet}, {Creevey}, {Davidson}, {De
  Ridder}, {Delchambre}, {Dell'Oro}, {Ducourant},
  {Fern{\'a}ndez-Hern{\'a}ndez}, {Fouesneau}, {Fr{\'e}mat}, {Galluccio},
  {Garc{\'\i}a-Torres}, {Gonz{\'a}lez-N{\'u}{\~n}ez}, {Gonz{\'a}lez-Vidal},
  {Gosset}, {Guy}, {Halbwachs}, {Hambly}, {Harrison}, {Hern{\'a}ndez},
  {Hestroffer}, {Hodgkin}, {Hutton}, {Jasniewicz}, {Jean-Antoine-Piccolo},
  {Jordan}, {Korn}, {Krone-Martins}, {Lanzafame}, {Lebzelter}, {L{\"o}ffler},
  {Manteiga}, {Marrese}, {Mart{\'\i}n-Fleitas}, {Moitinho}, {Mora}, {Muinonen},
  {Osinde}, {Pancino}, {Pauwels}, {Petit}, {Recio-Blanco}, {Richards},
  {Rimoldini}, {Robin}, {Sarro}, {Siopis}, {Smith}, {Sozzetti}, {S{\"u}veges},
  {Torra}, {van Reeven}, {Abbas}, {Abreu Aramburu}, {Accart}, {Aerts},
  {Altavilla}, {{\'A}lvarez}, {Alvarez}, {Alves}, {Anderson}, {Andrei},
  {Anglada Varela}, {Antiche}, {Antoja}, {Arcay}, {Astraatmadja}, {Bach},
  {Baker}, {Balaguer-N{\'u}{\~n}ez}, {Balm}, {Barache}, {Barata}, {Barbato},
  {Barblan}, {Barklem}, {Barrado}, {Barros}, {Barstow}, {Bartholom{\'e}
  Mu{\~n}oz}, {Bassilana}, {Becciani}, {Bellazzini}, {Berihuete}, {Bertone},
  {Bianchi}, {Bienaym{\'e}}, {Blanco-Cuaresma}, {Boch}, {Boeche}, {Bombrun},
  {Borrachero}, {Bossini}, {Bouquillon}, {Bourda}, {Bragaglia}, {Bramante},
  {Breddels}, {Bressan}, {Brouillet}, {Br{\"u}semeister}, {Brugaletta},
  {Bucciarelli}, {Burlacu}, {Busonero}, {Butkevich}, {Buzzi}, {Caffau},
  {Cancelliere}, {Cannizzaro}, {Cantat-Gaudin}, {Carballo}, {Carlucci},
  {Carrasco}, {Casamiquela}, {Castellani}, {Castro-Ginard}, {Charlot},
  {Chemin}, {Chiavassa}, {Cocozza}, {Costigan}, {Cowell}, {Crifo}, {Crosta},
  {Crowley}, {Cuypers}, {Dafonte}, {Damerdji}, {Dapergolas}, {David}, {David},
  {de Laverny}, {De Luise}, {De March}, {de Martino}, {de Souza}, {de Torres},
  {Debosscher}, {del Pozo}, {Delbo}, {Delgado}, {Delgado}, {Di Matteo},
  {Diakite}, {Diener}, {Distefano}, {Dolding}, {Drazinos}, {Dur{\'a}n},
  {Edvardsson}, {Enke}, {Eriksson}, {Esquej}, {Eynard Bontemps}, {Fabre},
  {Fabrizio}, {Faigler}, {Falc{\~a}o}, {Farr{\`a}s Casas}, {Federici},
  {Fedorets}, {Fernique}, {Figueras}, {Filippi}, {Findeisen}, {Fonti},
  {Fraile}, {Fraser}, {Fr{\'e}zouls}, {Gai}, {Galleti}, {Garabato},
  {Garc{\'\i}a-Sedano}, {Garofalo}, {Garralda}, {Gavel}, {Gavras}, {Gerssen},
  {Geyer}, {Giacobbe}, {Gilmore}, {Girona}, {Giuffrida}, {Glass}, {Gomes},
  {Granvik}, {Gueguen}, {Guerrier}, {Guiraud}, {Guti{\'e}rrez-S{\'a}nchez},
  {Haigron}, {Hatzidimitriou}, {Hauser}, {Haywood}, {Heiter}, {Helmi}, {Heu},
  {Hilger}, {Hobbs}, {Hofmann}, {Holland}, {Huckle}, {Hypki}, {Icardi},
  {Jan{\ss}en}, {Jevardat de Fombelle}, {Jonker}, {Juh{\'a}sz}, {Julbe},
  {Karampelas}, {Kewley}, {Klar}, {Kochoska}, {Kohley}, {Kolenberg},
  {Kontizas}, {Kontizas}, {Koposov}, {Kordopatis}, {Kostrzewa-Rutkowska},
  {Koubsky}, {Lambert}, {Lanza}, {Lasne}, {Lavigne}, {Le Fustec}, {Le
  Poncin-Lafitte}, {Lebreton}, {Leccia}, {Leclerc}, {Lecoeur-Taibi},
  {Lenhardt}, {Leroux}, {Liao}, {Licata}, {Lindstr{\o}m}, {Lister}, {Livanou},
  {Lobel}, {L{\'o}pez}, {Managau}, {Mann}, {Mantelet}, {Marchal}, {Marchant},
  {Marconi}, {Marinoni}, {Marschalk{\'o}}, {Marshall}, {Martino}, {Marton},
  {Mary}, {Massari}, {Matijevi{\v{c}}}, {Mazeh}, {McMillan}, {Messina},
  {Michalik}, {Millar}, {Molina}, {Molinaro}, {Moln{\'a}r}, {Montegriffo},
  {Mor}, {Morbidelli}, {Morel}, {Morris}, {Mulone}, {Muraveva}, {Musella},
  {Nelemans}, {Nicastro}, {Noval}, {O'Mullane}, {Ord{\'e}novic},
  {Ord{\'o}{\~n}ez-Blanco}, {Osborne}, {Pagani}, {Pagano}, {Pailler},
  {Palacin}, {Palaversa}, {Panahi}, {Pawlak}, {Piersimoni}, {Pineau}, {Plachy},
  {Plum}, {Poggio}, {Poujoulet}, {Pr{\v{s}}a}, {Pulone}, {Racero}, {Ragaini},
  {Rambaux}, {Ramos-Lerate}, {Regibo}, {Reyl{\'e}}, {Riclet}, {Ripepi}, {Riva},
  {Rivard}, {Rixon}, {Roegiers}, {Roelens}, {Romero-G{\'o}mez}, {Rowell},
  {Royer}, {Ruiz-Dern}, {Sadowski}, {Sagrist{\`a} Sell{\'e}s}, {Sahlmann},
  {Salgado}, {Salguero}, {Sanna}, {Santana-Ros}, {Sarasso}, {Savietto},
  {Schultheis}, {Sciacca}, {Segol}, {Segovia}, {S{\'e}gransan}, {Shih},
  {Siltala}, {Silva}, {Smart}, {Smith}, {Solano}, {Solitro}, {Sordo}, {Soria
  Nieto}, {Souchay}, {Spagna}, {Spoto}, {Stampa}, {Steele},
  {Steidelm{\"u}ller}, {Stephenson}, {Stoev}, {Suess}, {Surdej}, {Szabados},
  {Szegedi-Elek}, {Tapiador}, {Taris}, {Tauran}, {Taylor}, {Teixeira},
  {Terrett}, {Teyssand ier}, {Thuillot}, {Titarenko}, {Torra Clotet}, {Turon},
  {Ulla}, {Utrilla}, {Uzzi}, {Vaillant}, {Valentini}, {Valette}, {van Elteren},
  {Van Hemelryck}, {van Leeuwen}, {Vaschetto}, {Vecchiato}, {Veljanoski},
  {Viala}, {Vicente}, {Vogt}, {von Essen}, {Voss}, {Votruba}, {Voutsinas},
  {Walmsley}, {Weiler}, {Wertz}, {Wevers}, {Wyrzykowski}, {Yoldas},
  {{\v{Z}}erjal}, {Ziaeepour}, {Zorec}, {Zschocke}, {Zucker}, {Zurbach}, \&
  {Zwitter}}]{gaia18}
{Gaia Collaboration}, {Brown}, A.~G.~A., {Vallenari}, A., {et~al.} 2018, \aap,
  616, A1

\bibitem[{{Gallo} {et~al.}(2018){Gallo}, {Blue}, {Grupe}, {Komossa}, \&
  {Wilkins}}]{gallo18}
{Gallo}, L.~C., {Blue}, D.~M., {Grupe}, D., {Komossa}, S., \& {Wilkins}, D.~R.
  2018, \mnras, 478, 2557

\bibitem[{{Graham} {et~al.}(2019){Graham}, {Kulkarni}, {Bellm}, {Adams},
  {Barbarino}, {Blagorodnova}, {Bodewits}, {Bolin}, {Brady}, {Cenko}, {Chang},
  {Coughlin}, {De}, {Eadie}, {Farnham}, {Feindt}, {Franckowiak}, {Fremling},
  {Gezari}, {Ghosh}, {Goldstein}, {Golkhou}, {Goobar}, {Ho}, {Huppenkothen},
  {Ivezi{\'c}}, {Jones}, {Juric}, {Kaplan}, {Kasliwal}, {Kelley}, {Kupfer},
  {Lee}, {Lin}, {Lunnan}, {Mahabal}, {Miller}, {Ngeow}, {Nugent}, {Ofek},
  {Prince}, {Rauch}, {van Roestel}, {Schulze}, {Singer}, {Sollerman}, {Taddia},
  {Yan}, {Ye}, {Yu}, {Barlow}, {Bauer}, {Beck}, {Belicki}, {Biswas}, {Brinnel},
  {Brooke}, {Bue}, {Bulla}, {Burruss}, {Connolly}, {Cromer}, {Cunningham},
  {Dekany}, {Delacroix}, {Desai}, {Duev}, {Feeney}, {Flynn}, {Frederick},
  {Gal-Yam}, {Giomi}, {Groom}, {Hacopians}, {Hale}, {Helou}, {Henning},
  {Hover}, {Hillenbrand}, {Howell}, {Hung}, {Imel}, {Ip}, {Jackson}, {Kaspi},
  {Kaye}, {Kowalski}, {Kramer}, {Kuhn}, {Landry}, {Laher}, {Mao}, {Masci},
  {Monkewitz}, {Murphy}, {Nordin}, {Patterson}, {Penprase}, {Porter},
  {Rebbapragada}, {Reiley}, {Riddle}, {Rigault}, {Rodriguez}, {Rusholme}, {van
  Santen}, {Shupe}, {Smith}, {Soumagnac}, {Stein}, {Surace}, {Szkody}, {Terek},
  {Van Sistine}, {van Velzen}, {Vestrand}, {Walters}, {Ward}, {Zhang}, \&
  {Zolkower}}]{graham19}
{Graham}, M.~J., {Kulkarni}, S.~R., {Bellm}, E.~C., {et~al.} 2019, \pasp, 131,
  078001

\bibitem[{{Grupe} {et~al.}(2015){Grupe}, {Komossa}, \& {Saxton}}]{gruppe15}
{Grupe}, D., {Komossa}, S., \& {Saxton}, R. 2015, \apjl, 803, L28

\bibitem[{{Harrison} {et~al.}(2013){Harrison}, {Craig}, {Christensen},
  {Hailey}, \& {Zhang}}]{harrison13}
{Harrison}, F.~A., {Craig}, W.~W., {Christensen}, F.~E., {Hailey}, C.~J., \&
  {Zhang}, W.~W. 2013, \apj, 770, 103

\bibitem[{{HI4PI Collaboration} {et~al.}(2016){HI4PI Collaboration}, {Ben
  Bekhti}, {Fl{\"o}er}, {Keller}, {Kerp}, {Lenz}, {Winkel}, {Bailin},
  {Calabretta}, {Dedes}, {Ford}, {Gibson}, {Haud}, {Janowiecki}, {Kalberla},
  {Lockman}, {McClure-Griffiths}, {Murphy}, {Nakanishi}, {Pisano}, \&
  {Staveley-Smith}}]{HI4PI16}
{HI4PI Collaboration}, {Ben Bekhti}, N., {Fl{\"o}er}, L., {et~al.} 2016, \aap,
  594, A116

\bibitem[{{Ho} {et~al.}(2020){Ho}, {Kulkarni}, {Perley}, {Cenko}, {Corsi},
  {Schulze}, {Lunnan}, {Sollerman}, {Gal-Yam}, {Anand}, {Barbarino}, {Bellm},
  {Bruch}, {Burns}, {De}, {Dekany}, {Delacroix}, {Duev}, {Fremling},
  {Goldstein}, {Golkhou}, {Graham}, {Hale}, {Kasliwal}, {Kupfer}, {Laher},
  {Martikainen}, {Masci}, {Neill}, {Rusholme}, {Shupe}, {Soumagnac},
  {Strotjohann}, {Taggart}, {Tartaglia}, {Yan}, \& {Zolkower}}]{ho20}
{Ho}, A.~Y.~Q., {Kulkarni}, S.~R., {Perley}, D.~A., {et~al.} 2020, arXiv
  e-prints, arXiv:2004.10406

\bibitem[{{Holoien} {et~al.}(2016){Holoien}, {Kochanek}, {Prieto}, {Stanek},
  {Dong}, {Shappee}, {Grupe}, {Brown}, {Basu}, {Beacom}, {Bersier},
  {Brimacombe}, {Danilet}, {Falco}, {Guo}, {Jose}, {Herczeg}, {Long},
  {Pojmanski}, {Simonian}, {Szczygie{\l}}, {Thompson}, {Thorstensen}, {Wagner},
  \& {Wo{\'z}niak}}]{holoien16}
{Holoien}, T.~W.~S., {Kochanek}, C.~S., {Prieto}, J.~L., {et~al.} 2016, \mnras,
  455, 2918

\bibitem[{{Holoien} {et~al.}(2019){Holoien}, {Vallely}, {Auchettl}, {Stanek},
  {Kochanek}, {French}, {Prieto}, {Shappee}, {Brown}, {Fausnaugh}, {Dong},
  {Thompson}, {Bose}, {Neustadt}, {Cacella}, {Brimacombe}, {Kendurkar},
  {Beaton}, {Boutsia}, {Chomiuk}, {Connor}, {Morrell}, {Newman}, {Rudie},
  {Shishkovksy}, \& {Strader}}]{holoien19}
{Holoien}, T. W.~S., {Vallely}, P.~J., {Auchettl}, K., {et~al.} 2019, \apj,
  883, 111

\bibitem[{{Kauffmann} {et~al.}(2003){Kauffmann}, {Heckman}, {Tremonti},
  {Brinchmann}, {Charlot}, {White}, {Ridgway}, {Brinkmann}, {Fukugita}, {Hall},
  {Ivezi{\'c}}, {Richards}, \& {Schneider}}]{kauffmann03}
{Kauffmann}, G., {Heckman}, T.~M., {Tremonti}, C., {et~al.} 2003, \mnras, 346,
  1055

\bibitem[{{Kewley} {et~al.}(2001){Kewley}, {Heisler}, {Dopita}, \&
  {Lumsden}}]{kewley01}
{Kewley}, L.~J., {Heisler}, C.~A., {Dopita}, M.~A., \& {Lumsden}, S. 2001,
  ApJS, 132, 37

\bibitem[{{Khabibullin} {et~al.}(2020){Khabibullin}, {Sunyaev}, {Churazov},
  {Gilfanov}, {Medvedev}, \& {Sazonov}}]{khabibullin20}
{Khabibullin}, I., {Sunyaev}, R., {Churazov}, E., {et~al.} 2020, The
  Astronomer's Telegram, 13494, 1

\bibitem[{{Komossa}(2015)}]{komossa15}
{Komossa}, S. 2015, Journal of High Energy Astrophysics, 7, 148

\bibitem[{{Lang} {et~al.}(2016){Lang}, {Hogg}, \& {Mykytyn}}]{lang16}
{Lang}, D., {Hogg}, D.~W., \& {Mykytyn}, D. 2016, {The Tractor: Probabilistic
  astronomical source detection and measurement}, ascl:1604.008

\bibitem[{{Law-Smith} {et~al.}(2017){Law-Smith}, {Ramirez-Ruiz}, {Ellison}, \&
  {Foley}}]{lawsmith17}
{Law-Smith}, J., {Ramirez-Ruiz}, E., {Ellison}, S.~L., \& {Foley}, R.~J. 2017,
  \apj, 850, 22

\bibitem[{{Liu} {et~al.}(2014){Liu}, {Li}, \& {Komossa}}]{liu14_tde}
{Liu}, F.~K., {Li}, S., \& {Komossa}, S. 2014, \apj, 786, 103

\bibitem[{{Lusso} \& {Risaliti}(2016)}]{lusso16}
{Lusso}, E., \& {Risaliti}, G. 2016, \apj, 819, 154

\bibitem[{{Lusso} {et~al.}(2010){Lusso}, {Comastri}, {Vignali}, {Zamorani},
  {Brusa}, {Gilli}, {Iwasawa}, {Salvato}, {Civano}, {Elvis}, {Merloni},
  {Bongiorno}, {Trump}, {Koekemoer}, {Schinnerer}, {Le Floc'h}, \&
  {Cappelluti}}]{lusso10}
{Lusso}, E., {Comastri}, A., {Vignali}, C., {et~al.} 2010, \aap, 512, A34+

\bibitem[{{Madsen} {et~al.}(2015){Madsen}, {Harrison}, {Markwardt}, {An},
  {Grefenstette}, {Bachetti}, {Miyasaka}, {Kitaguchi}, {Bhalerao}, {Boggs},
  {Christensen}, {Craig}, {Forster}, {Fuerst}, {Hailey}, {Perri}, {Puccetti},
  {Rana}, {Stern}, {Walton}, {J{\o}rgen Westergaard}, \& {Zhang}}]{madsen15}
{Madsen}, K.~K., {Harrison}, F.~A., {Markwardt}, C.~B., {et~al.} 2015, \apjs,
  220, 8

\bibitem[{{Mainzer} {et~al.}(2011){Mainzer}, {Bauer}, {Grav}, {Masiero},
  {Cutri}, {Dailey}, {Eisenhardt}, {McMillan}, {Wright}, {Walker}, {Jedicke},
  {Spahr}, {Tholen}, {Alles}, {Beck}, {Brand enburg}, {Conrow}, {Evans},
  {Fowler}, {Jarrett}, {Marsh}, {Masci}, {McCallon}, {Wheelock}, {Wittman},
  {Wyatt}, {DeBaun}, {Elliott}, {Elsbury}, {Gautier}, {Gomillion}, {Leisawitz},
  {Maleszewski}, {Micheli}, \& {Wilkins}}]{mainzer11}
{Mainzer}, A., {Bauer}, J., {Grav}, T., {et~al.} 2011, \apj, 731, 53

\bibitem[{{Makishima} {et~al.}(1986){Makishima}, {Maejima}, {Mitsuda}, {Bradt},
  {Remillard}, {Tuohy}, {Hoshi}, \& {Nakagawa}}]{makishima86}
{Makishima}, K., {Maejima}, Y., {Mitsuda}, K., {et~al.} 1986, \apj, 308, 635

\bibitem[{{Masci} {et~al.}(2019){Masci}, {Laher}, {Rusholme}, {Shupe}, {Groom},
  {Surace}, {Jackson}, {Monkewitz}, {Beck}, {Flynn}, {Terek}, {Landry},
  {Hacopians}, {Desai}, {Howell}, {Brooke}, {Imel}, {Wachter}, {Ye}, {Lin},
  {Cenko}, {Cunningham}, {Rebbapragada}, {Bue}, {Miller}, {Mahabal}, {Bellm},
  {Patterson}, {Juri{\'c}}, {Golkhou}, {Ofek}, {Walters}, {Graham}, {Kasliwal},
  {Dekany}, {Kupfer}, {Burdge}, {Cannella}, {Barlow}, {Van Sistine}, {Giomi},
  {Fremling}, {Blagorodnova}, {Levitan}, {Riddle}, {Smith}, {Helou}, {Prince},
  \& {Kulkarni}}]{masci19}
{Masci}, F.~J., {Laher}, R.~R., {Rusholme}, B., {et~al.} 2019, \pasp, 131,
  018003

\bibitem[{{Mattila} {et~al.}(2018){Mattila}, {P{\'e}rez-Torres}, {Efstathiou},
  {Mimica}, {Fraser}, {Kankare}, {Alberdi}, {Aloy}, {Heikkil{\"a}}, {Jonker},
  {Lundqvist}, {Mart{\'\i}-Vidal}, {Meikle}, {Romero-Ca{\~n}izales}, {Smartt},
  {Tsygankov}, {Varenius}, {Alonso-Herrero}, {Bondi}, {Fransson},
  {Herrero-Illana}, {Kangas}, {Kotak}, {Ram{\'\i}rez-Olivencia},
  {V{\"a}is{\"a}nen}, {Beswick}, {Clements}, {Greimel}, {Harmanen},
  {Kotilainen}, {Nandra}, {Reynolds}, {Ryder}, {Walton}, {Wiik}, \&
  {{\"O}stlin}}]{mattila18}
{Mattila}, S., {P{\'e}rez-Torres}, M., {Efstathiou}, A., {et~al.} 2018,
  Science, 361, 482

\bibitem[{{McMullin} {et~al.}(2007){McMullin}, {Waters}, {Schiebel}, {Young},
  \& {Golap}}]{mcmullin07}
{McMullin}, J.~P., {Waters}, B., {Schiebel}, D., {Young}, W., \& {Golap}, K.
  2007, in Astronomical Society of the Pacific Conference Series, Vol. 376,
  Astronomical Data Analysis Software and Systems XVI, ed. R.~A. {Shaw},
  F.~{Hill}, \& D.~J. {Bell}, 127

\bibitem[{{Merloni} {et~al.}(2012){Merloni}, {Predehl}, {Becker},
  {B{\"o}hringer}, {Boller}, {Brunner}, {Brusa}, {Dennerl}, {Freyberg},
  {Friedrich}, {Georgakakis}, {Haberl}, {Hasinger}, {Meidinger}, {Mohr},
  {Nandra}, {Rau}, {Reiprich}, {Robrade}, {Salvato}, {Santangelo}, {Sasaki},
  {Schwope}, {Wilms}, \& {German eROSITA Consortium}}]{merloni12}
{Merloni}, A., {Predehl}, P., {Becker}, W., {et~al.} 2012, ArXiv e-prints,
  arXiv:1209.3114

\bibitem[{{Merloni} {et~al.}(2015){Merloni}, {Dwelly}, {Salvato},
  {Georgakakis}, {Greiner}, {Krumpe}, {Nandra}, {Ponti}, \& {Rau}}]{merloni15}
{Merloni}, A., {Dwelly}, T., {Salvato}, M., {et~al.} 2015, \mnras, 452, 69

\bibitem[{{Mitsuda} {et~al.}(1984){Mitsuda}, {Inoue}, {Koyama}, {Makishima},
  {Matsuoka}, {Ogawara}, {Shibazaki}, {Suzuki}, {Tanaka}, \&
  {Hirano}}]{mitsuda84}
{Mitsuda}, K., {Inoue}, H., {Koyama}, K., {et~al.} 1984, \pasj, 36, 741

\bibitem[{{Oke} {et~al.}(1995){Oke}, {Cohen}, {Carr}, {Cromer}, {Dingizian},
  {Harris}, {Labrecque}, {Lucinio}, {Schaal}, {Epps}, \& {Miller}}]{oke95}
{Oke}, J.~B., {Cohen}, J.~G., {Carr}, M., {et~al.} 1995, \pasp, 107, 375

\bibitem[{{Perley}(2019)}]{perley19}
{Perley}, D.~A. 2019, \pasp, 131, 084503

\bibitem[{{Pounds} {et~al.}(1995){Pounds}, {Done}, \& {Osborne}}]{pounds95}
{Pounds}, K.~A., {Done}, C., \& {Osborne}, J.~P. 1995, \mnras, 277, L5

\bibitem[{{Rees}(1988)}]{rees88}
{Rees}, M.~J. 1988, \nat, 333, 523

\bibitem[{{Ricci} {et~al.}(2017){Ricci}, {Trakhtenbrot}, {Koss}, {Ueda}, {Del
  Vecchio}, {Treister}, {Schawinski}, {Paltani}, {Oh}, {Lamperti}, {Berney},
  {Gand hi}, {Ichikawa}, {Bauer}, {Ho}, {Asmus}, {Beckmann}, {Soldi},
  {Balokovi{\'c}}, {Gehrels}, \& {Markwardt}}]{ricci17b}
{Ricci}, C., {Trakhtenbrot}, B., {Koss}, M.~J., {et~al.} 2017, \apjs, 233, 17

\bibitem[{{Ricci} {et~al.}(2020){Ricci}, {Kara}, {Loewenstein}, {Trakhtenbrot},
  {Arcavi}, {Remillard}, {Fabian}, {Gendreau}, {Arzoumanian}, {Li}, {Ho},
  {MacLeod}, {Cackett}, {Altamirano}, {Gand hi}, {Kosec}, {Pasham}, {Steiner},
  \& {Chan}}]{ricci20}
{Ricci}, C., {Kara}, E., {Loewenstein}, M., {et~al.} 2020, \apjl, 898, L1

\bibitem[{{Risaliti} {et~al.}(2009){Risaliti}, {Salvati}, {Elvis}, {Fabbiano},
  {Baldi}, {Bianchi}, {Braito}, {Guainazzi}, {Matt}, {Miniutti}, {Reeves},
  {Soria}, \& {Zezas}}]{risaliti09}
{Risaliti}, G., {Salvati}, M., {Elvis}, M., {et~al.} 2009, \mnras, 393, L1

\bibitem[{{Saxton} {et~al.}(2008){Saxton}, {Read}, {Esquej}, {Freyberg},
  {Altieri}, \& {Bermejo}}]{saxton12a}
{Saxton}, R.~D., {Read}, A.~M., {Esquej}, P., {et~al.} 2008, \aap, 480, 611

\bibitem[{{Saxton} {et~al.}(2012){Saxton}, {Read}, {Esquej}, {Komossa},
  {Dougherty}, {Rodriguez-Pascual}, \& {Barrado}}]{saxton12}
---. 2012, \aap, 541, A106

\bibitem[{{Saxton} {et~al.}(2017){Saxton}, {Read}, {Komossa}, {Lira},
  {Alexander}, \& {Wieringa}}]{saxton17}
{Saxton}, R.~D., {Read}, A.~M., {Komossa}, S., {et~al.} 2017, \aap, 598, A29

\bibitem[{{Shemmer} {et~al.}(2006){Shemmer}, {Brandt}, {Netzer}, {Maiolino}, \&
  {Kaspi}}]{shemmer06}
{Shemmer}, O., {Brandt}, W.~N., {Netzer}, H., {Maiolino}, R., \& {Kaspi}, S.
  2006, \apjl, 646, L29

\bibitem[{{Shemmer} {et~al.}(2008){Shemmer}, {Brandt}, {Netzer}, {Maiolino}, \&
  {Kaspi}}]{shemmer08}
---. 2008, \apj, 682, 81

\bibitem[{{Sobolewska} \& {Papadakis}(2009)}]{sobolewska09}
{Sobolewska}, M.~A., \& {Papadakis}, I.~E. 2009, \mnras, 399, 1597

\bibitem[{{Soderberg} {et~al.}(2009){Soderberg}, {Grindlay}, {Bloom}, {Gezari},
  {Piro}, {Belloni}, {Liu}, {Berger}, {Coppi}, {Kawai}, {Gehrels}, {Metzger},
  {Allen}, {Barret}, {Bazzano}, {Bignami}, {Caraveo}, {Corbel}, {De Luca},
  {Fabbiano}, {Finger}, {Feroci}, {Hartmann}, {Hong}, {Jernigan}, {Kaaret},
  {Kouveliotou}, {Kutrev}, {Loeb}, {Paizis}, {Pareschi}, {Skinner}, {Di
  Stefano}, {Tagliaferri}, {Ubertini}, {van der Klis}, \&
  {Wilson-Hodge}}]{soderberg09}
{Soderberg}, A., {Grindlay}, J.~E., {Bloom}, J.~S., {et~al.} 2009, in
  astro2010: The Astronomy and Astrophysics Decadal Survey, Vol. 2010, 278

\bibitem[{{Steffen} {et~al.}(2006){Steffen}, {Strateva}, {Brandt}, {Alexander},
  {Koekemoer}, {Lehmer}, {Schneider}, \& {Vignali}}]{steffen06}
{Steffen}, A.~T., {Strateva}, I., {Brandt}, W.~N., {et~al.} 2006, \aj, 131,
  2826

\bibitem[{{Steiner} {et~al.}(2009){Steiner}, {Narayan}, {McClintock}, \&
  {Ebisawa}}]{steiner09}
{Steiner}, J.~F., {Narayan}, R., {McClintock}, J.~E., \& {Ebisawa}, K. 2009,
  \pasp, 121, 1279

\bibitem[{{Stern} {et~al.}(2012){Stern}, {Assef}, {Benford}, {Blain}, {Cutri},
  {Dey}, {Eisenhardt}, {Griffith}, {Jarrett}, {Lake}, {Masci}, {Petty},
  {Stanford}, {Tsai}, {Wright}, {Yan}, {Harrison}, \& {Madsen}}]{stern12}
{Stern}, D., {Assef}, R.~J., {Benford}, D.~J., {et~al.} 2012, \apj, 753, 30

\bibitem[{{Stone} \& {Metzger}(2016)}]{stone16}
{Stone}, N.~C., \& {Metzger}, B.~D. 2016, \mnras, 455, 859

\bibitem[{{Strubbe} \& {Quataert}(2009)}]{strubbe09}
{Strubbe}, L.~E., \& {Quataert}, E. 2009, \mnras, 400, 2070

\bibitem[{{Trakhtenbrot} {et~al.}(2017){Trakhtenbrot}, {Ricci}, {Koss},
  {Schawinski}, {Mushotzky}, {Ueda}, {Veilleux}, {Lamperti}, {Oh}, {Treister},
  {Stern}, {Harrison}, {Balokovi{\'c}}, \& {Gehrels}}]{trakhtenbrot17}
{Trakhtenbrot}, B., {Ricci}, C., {Koss}, M.~J., {et~al.} 2017, \mnras, 470, 800

\bibitem[{{Tremonti} {et~al.}(2004){Tremonti}, {Heckman}, {Kauffmann},
  {Brinchmann}, {Charlot}, {White}, {Seibert}, {Peng}, {Schlegel}, {Uomoto},
  {Fukugita}, \& {Brinkmann}}]{tremonti04}
{Tremonti}, C.~A., {Heckman}, T.~M., {Kauffmann}, G., {et~al.} 2004, \apj, 613,
  898

\bibitem[{{van Velzen} {et~al.}(2016){van Velzen}, {Anderson}, {Stone},
  {Fraser}, {Wevers}, {Metzger}, {Jonker}, {van der Horst}, {Staley}, {Mendez},
  {Miller-Jones}, {Hodgkin}, {Campbell}, \& {Fender}}]{vanvelzen16}
{van Velzen}, S., {Anderson}, G.~E., {Stone}, N.~C., {et~al.} 2016, Science,
  351, 62

\bibitem[{{van Velzen} {et~al.}(2019){van Velzen}, {Gezari}, {Cenko}, {Kara},
  {Miller-Jones}, {Hung}, {Bright}, {Roth}, {Blagorodnova}, {Huppenkothen},
  {Yan}, {Ofek}, {Sollerman}, {Frederick}, {Ward}, {Graham}, {Fender},
  {Kasliwal}, {Canella}, {Stein}, {Giomi}, {Brinnel}, {van Santen}, {Nordin},
  {Bellm}, {Dekany}, {Fremling}, {Golkhou}, {Kupfer}, {Kulkarni}, {Laher},
  {Mahabal}, {Masci}, {Miller}, {Neill}, {Riddle}, {Rigault}, {Rusholme},
  {Soumagnac}, \& {Tachibana}}]{vanvelzen19}
{van Velzen}, S., {Gezari}, S., {Cenko}, S.~B., {et~al.} 2019, \apj, 872, 198

\bibitem[{{van Velzen} {et~al.}(2020){van Velzen}, {Gezari}, {Hammerstein},
  {Roth}, {Frederick}, {Ward}, {Hung}, {Cenko}, {Stein}, {Perley}, {Taggart},
  {Sollerman}, {Andreoni}, {Bellm}, {Brinnel}, {De}, {Dekany}, {Feeney},
  {Foley}, {Fremling}, {Giomi}, {Golkhou}, {Ho}, {Kasliwal}, {Kilpatrick},
  {Kulkarni}, {Kupfer}, {Laher}, {Mahabal}, {Masci}, {Nordin}, {Riddle},
  {Rusholme}, {Sharma}, {van Santen}, {Shupe}, \& {Soumagnac}}]{vanvelzen20}
{van Velzen}, S., {Gezari}, S., {Hammerstein}, E., {et~al.} 2020, arXiv
  e-prints, arXiv:2001.01409

\bibitem[{{Voges} {et~al.}(1999){Voges}, {Aschenbach}, {Boller},
  {Br{\"a}uninger}, {Briel}, {Burkert}, {Dennerl}, {Englhauser}, {Gruber},
  {Haberl}, {Hartner}, {Hasinger}, {K{\"u}rster}, {Pfeffermann}, {Pietsch},
  {Predehl}, {Rosso}, {Schmitt}, {Tr{\"u}mper}, \& {Zimmermann}}]{voges99}
{Voges}, W., {Aschenbach}, B., {Boller}, T., {et~al.} 1999, \aap, 349, 389

\bibitem[{{Weisskopf}(1999)}]{weisskopf99}
{Weisskopf}, M.~C. 1999, ArXiv Astrophysics e-prints, astro-ph/9912097

\bibitem[{{Wevers} {et~al.}(2017){Wevers}, {van Velzen}, {Jonker}, {Stone},
  {Hung}, {Onori}, {Gezari}, \& {Blagorodnova}}]{wevers17}
{Wevers}, T., {van Velzen}, S., {Jonker}, P.~G., {et~al.} 2017, \mnras, 471,
  1694

\bibitem[{{Wevers} {et~al.}(2019){Wevers}, {Stone}, {van Velzen}, {Jonker},
  {Hung}, {Auchettl}, {Gezari}, {Onori}, {Mata S{\'a}nchez},
  {Kostrzewa-Rutkowska}, \& {Casares}}]{wevers19}
{Wevers}, T., {Stone}, N.~C., {van Velzen}, S., {et~al.} 2019, \mnras, 487,
  4136

\bibitem[{{Woo} {et~al.}(2013){Woo}, {Schulze}, {Park}, {Kang}, {Kim}, \&
  {Riechers}}]{woo13}
{Woo}, J.-H., {Schulze}, A., {Park}, D., {et~al.} 2013, \apj, 772, 49

\end{thebibliography}

\end{document}